\documentclass{aa}
\usepackage{graphicx}
\usepackage{aalongtable}
\usepackage{natbib}
\bibpunct{(}{)}{;}{a}{}{,}
\begin{document}
\title{Radio-optical scrutiny of compact AGN: Correlations between properties of pc-scale jets and optical nuclear emission}


   \titlerunning{Correlations between radio-optical properties of compact AGN}

   \authorrunning{Arshakian et al.}

   \author{T.G. Arshakian\inst{1},
      J. Torrealba\inst{2},
      V.H. Chavushyan\inst{3},
          E. Ros\inst{1,4},
          M.L. Lister\inst{5},
          I. Cruz-Gonz\'alez\inst{2},	
     \and
     J.A. Zensus\inst{1}
          }

   \offprints{T.G. Arshakian}

   \institute{
   Max-Planck-Institut f\"ur Radioastronomie, Auf dem H\"ugel 69,
   53121 Bonn, Germany\\
   \email{tigar@mpifr-bonn.mpg.de}
         \and
   Instituto de Astronom\'{\i}a, Universidad Nacional Aut\'onoma de M\'exico, Apartado Postal 70-264,
   04510 M\'exico D.F., M\'exico
       \and
   Instituto Nacional de Astrof\'{\i}sica \'Optica y
   Electr\'onica, Apartado Postal 51 y 216, 72000 Puebla, Pue, M\'exico
       \and
   Departament d'Astronomia i Astrof\'{\i}sica, Universitat de Val\`encia, E-46100 Burjassot, Spain
       \and
   Department of Physics, Purdue University, 525 Northwestern Avenue,
   West Lafayette, IN 47907, USA
             }

   \date{Received <date> / Accepted <date>}


  \abstract
   {}
   {We study the correlations between the VLBA radio emission at 15 GHz, extended emission at 151 MHz, and optical nuclear emission at 5100\,\AA\ for a complete sample of 135 compact jets.}
   {We use the partial Kendall's tau correlation analysis to check the link between radio properties of parsec-scale jets and optical luminosities of host active galactic nuclei (AGN).}
   {We find a significant positive correlation for 99 quasars between optical nuclear luminosities and total radio (VLBA) luminosities of unresolved cores at 15\,GHz originated at milliarcseconds scales. For 18 BL Lacs, the optical continuum emission correlates with the radio emission of the jet at 15\,GHz. We suggest that the radio and optical emission are beamed and originate in the innermost part of the sub--parsec-scale jet in quasars. Analysis of the relation between the apparent speed of the jet and the optical nuclear luminosity at 5100\,\AA\ supports the relativistic beaming model for the optical emission generated in the jet, and allows the peak values of the intrinsic optical luminosity of the jet and its Lorentz factor to be estimated for the populations of quasars and radio galaxies. The radio-loudness of quasars  is found to increase at high redshifts, which can be a result of lower efficiency of the accretion in AGN having higher radio luminosities. A strong positive correlation is found between the intrinsic kinetic power of the jet and the apparent luminosities of the total and the unresolved core emission of the jet at 15\,GHz. This correlation is interpreted in terms of intrinsically more luminous parsec-scale jet producing more luminous extended structure which is detectable at low radio frequencies, 151\,MHz. A possibility that the low frequency radio emission is relativistically beamed in superluminal AGN and therefore correlates with radio luminosity of the jet at 15\,GHz can not be ruled out (abridged).}
   {}

   \maketitle
%

\section{Introduction}

The orientation-based unification schemes of radio-loud active galactic nuclei (AGN)
\citep{barthel89,urry95} suggest that the continuum and broad-line
emission from their central engine are seen directly in powerful FR II
\citep{fanaroff74} quasars, while in FR II radio galaxies the central
emission can be partially/completely hidden by obscuring material
(``dusty torus''). In this scheme the presence of relativistic jets
implies that radio-loud quasars are the relativistically beamed
counterparts of radio galaxies. The beamed synchrotron emission from
the base of the jet may extend to optical wavelengths in quasars
\citep{impey90} and even in the relatively unbeamed radio galaxies
\citep{chiaberge02,hardcastle00}. We should expect that in quasars the
emission from both the central engine (thermal?) and the jet
(non-thermal) contribute to the total power, whereas in radio galaxies
the bulk of the optical continuum emission may be attributed to the
relativistic jet rather than the central engine hidden by the torus \citep[e.g.,][]{arshakian10,tavares10}.

One approach to investigate the physical processes in active galactic nuclei (AGN) at scales not
reachable by present-day telescopes is to study the correlations
between radiative energy in different wavebands. 
There is evidence that the beamed synchrotron emission from the base of the radio jet extends to visible
wavelengths in BL Lacs and quasars \citep{wills92}. Several authors \citep{hardcastle00,chiaberge02,kharb04} investigated the correlations
between the radio and optical regimes for radio-loud FRI-FRII radio galaxies with the sample size $\la 65$ objects. On the basis of the correlations between unresolved optical core emission (in the high-resolution images with the \emph{Hubble Space Telescope}) and the VLA (Very Large Array) radio core emission of the jet on scales of milliarcseconds \citep[see][]{hardcastle00}, and color
information, these authors argued that optical nuclei are due to synchrotron radiation from the jet. The VLBI (Very Long Baseline Interferometry) imaging of jets at 15\,GHz reaches an unprecedented milliarcsecond resolution \citep{ken98,zensus02,ken04} and the VLBI cores are resolved on submilliarcsecond scales \citep{kovalev05}. The VLBA provides a high reliability in the results due to the excellent calibration properties and the repeatability of observations. Here we investigate the radio-optical correlation between the VLBA core emission at 15\,GHz and the optical nuclear emission at 5100\,\AA\ to test a single production mechanism for radio and optical continuum emission on scales of submilliarcseconds. 

The relativistic outflows of plasma material form
near the central nucleus and trace the pc-scale broad-line region
and kpc-scale narrow-line region transporting the kinetic energy
preserved in the jet to hundreds of kiloparsecs away from the central engine. To test the link between the properties of the pc-scale jet and extended radio structure on kiloparsec scales, as well as the correlation between VLBI radio and optical nuclear luminosities  we use the complete sample of 135 core-dominated AGN possessing relativistic jets \citep{lister05}.

In Sect.~\ref{sec:rsample} we introduce the radio sample of AGN, and in Sect.~\ref{sec:jetPar} we define the radio parameters of their compact jets. The optical nuclear luminosities are derived in Sect.~\ref{sec:optLum}. Correlations between radio properties of the jet, and between radio properties and optical nuclear luminosities are discussed in Sect.~\ref{sec:rCorr} and \ref{sec:roCorr} respectively. 


Throughout the paper a flat cosmology model is used with
$\Omega_{m}=0.3$ ($\Omega_{\Lambda}+\Omega_{m}=1$) and $H_0=70$
km\,s$^{-1}$\,Mpc$^{-1}$.

\section{The sample of compact AGN}
\label{sec:rsample}
To analyze the correlations between optical continuum emission and the
characteristics of radio jets on sub-parsec scales we use the complete statistical sample of compact AGN sources 
\citep{lister05} and the sample of 250 compact extragalactic radio sources at 15\,GHz compiled by
\citet{kovalev05}. The samples consist of compact AGN from the
MOJAVE (Monitoring of Jets in AGN with VLBA Experiments) program
and the VLBA 15\,GHz monitoring survey. 
The MOJAVE-1 sample is a flux-density limited complete sample of 135
AGN with redshifts reaching up to 3.4. Most of the MOJAVE-1 AGN have
flat spectra and their total flux density at 15\,GHz is greater than
1.5\,Jy for sources in the Northern hemisphere ($\delta>0^{\circ}$)
and $>2$\,Jy in the Southern hemisphere
($-20^{\circ}<\delta<0^{\circ}$). Of the 135 radio-loud AGN in
the MOJAVE-1 sample there are 101 quasars, 22 BL Lacs, 8
galaxies, and 3 unidentified objects (classification of AGN is adopted from \citet{veron03}). 
Based on the extended emission at 1.4\,GHz, \cite{kharb10} showed that the radio luminosities of a substantial fraction of the MOJAVE quasars fall in the intermediate power range characteristic  for FR\,I and FR\,II radio structures. Radio luminosities and morphologies of BL Lacs at 1.4 GHz are consistent with the FR\,II morphology and radio power of quasars.

This complete sample is used to infer relationships between optical luminosity and 
properties of compact jet. These relations are also checked for the full sample of 250
AGN, which includes additional compact AGN of special interest
that did not meet the criteria of the complete sample.
The full sample of 250 AGN includes 188 quasars, 36 BL Lacs, 20 radio
galaxies, and 6 sources with no optical identification.  \\


\emph{Biases of the sample}. According to the relativistic beaming model, the intrinsic flux density of the relativistic jet is enhanced toward the jet direction by relativistic Doppler beaming effect. The beamed emission appears to be stronger at small viewing angles of the jet \citep{pearson87,cohen07}. Compact jets with high intrinsic radio luminosities  can be seen at larger viewing angles, while the intrinsically faint jets fade away at larger viewing angles and are not included in the sample. In combination with the jet speed, this effect introduces a bias towards fast and bright jets to be oriented near to the line of sight.  At small angles we should have a wider range of Lorentz factors and intrinsic radio luminosities, while at large angles only slow and intrinsically bright jets are included. Low-redshift, bright sources can also have large viewing angles evidenced by two-sided compact jets. 

The sample of compact radio jets compiled by a flux limit has various selection effects, which arise from a combination of intrinsic luminosity, luminosity distance, and Doppler factor. Monte Carlo simulations of flux-limited, beamed samples suggest that the mean Doppler factor does not generally increase with redshift \citep{lister97}.  For low redshifts ($z<0.6$),
there is a deficit of high-Doppler factor, low viewing angle sources since they are rare in the population; the volume element at low-$z$ is small, and these sources are close enough that they do not need to be as beamed to be in included the flux-limited sample. Basically beaming is not as important a factor at lower redshifts in determining the makeup of a flux-limited sample as at high redshifts, $z>0.6$, according to \cite{lister97}. These authors showed that the viewing angle of the jet has a wide range at low redshifts, while at $z>0.6$ the mean viewing angle is pretty much constant, as was shown by simulations. Essentially, there is enough range in the intrinsic luminosity function, and the volume is large enough, so that at $z>0.6$, a combination of slightly less aligned sources with higher intrinsic powers, and vice versa, are included in the flux-limited sample.

The variability Doppler factors (measured from the variability timescales of radio flares at different frequencies; see Hovatta et al. 2009) of about 50\% of the MOJAVE blazars were estimated with the total flux density observations obtained with the Mets\"ahovi single-dish 14-meter telescope at 22\,GHz and 37\,GHz. These observations allow the measurements of variability Doppler factors for all flaring jets to be made, while for jets too compact to be observed with the VLBA at 15\,GHz, the apparent speed cannot be measured due to the absence of moving jet components. Around 50\% of the MOJAVE-1 core-dominated compact jets have no measurements of Doppler factors \citep{hovatta09}, because 53 sources were not monitored or analyzed, and 12 AGN had no detection of moving components (6 BL Lacs and 6 quasars). The jets of later sources are believed to have very small viewing angles. The selection of AGN with known apparent speeds thus should introduce a bias towards larger viewing angles, which can be significant for the sample of 16 BL Lacs with known apparent speeds. 

Another issue is the accurate division between BL Lacs and quasars. \cite{kharb10} showed that the division of blazars into two subclasses by an emission-line equivalent width of 5\,\AA\ is arbitrary in terms of radio jet properties of AGN forming the MOJAVE sample. Some correlation tests for BL Lacs and quasars differ significantly when two subclasses are considered separately, indicating that it is likely that the population of BL Lacs is inhomogeneous and consists of blazars with both FR\,I and FR\,II morphological types, while the population of quasars includes only FR\,II radio galaxies. Other problems relating to the precise delineation between BL Lacs and quasars is discussed in \cite{kovalev05}. Because we aim to study the radio-optical correlations for a relatively small MOJAVE-1 sample of 135 AGN, and find significant correlations for BL Lacs, quasars and radio galaxies, we adopt the original classification of blazars provided by \cite{veron03} to keep the subsamples under investigation large enough to find significant correlations.

\section{Parameters of the jet}
\label{sec:jetPar}
The relativistic plasma outflows oriented near to the line of sight are observed as one-sided radio core-jet structures (as seen on VLBA images) due to relativistic brightening of the radio emission from the approaching jet, while outflows oriented at large angles present a two-sided jet structure. The apparent speed of the fastest jet component is assumed to be equal to the speed of the jet flow.
Kovalev et al. (2005) studied the central regions of 250 compact AGN at 15\,GHz. They derived the total flux densities from each VLBA image ($S_{\rm VLBA}$) and the flux densities from unresolved, most compact regions of the VLBA cores ($S_{\rm un}$). The most compact structures extend at sub-milliarcsecond scales (0.02-0.06 mas). The uncertainty in $S_{\rm VLBA}$ and $S_{\rm un}$ is about 5\,\%. The flux density from the extended jet was defined as $S_{\rm jet}=S_{\rm VLBA}-S_{\rm core}$, where $S_{\rm core}$ is the VLBA resolved core flux (Kovalev et al. 2005). The luminosity of the VLBA component, unresolved core, and jet components are
\begin{equation}
\label{eq:lum}
L = 4\pi S d_{\rm L}^2 (1+z)^{-(1+\alpha)},
\end{equation}
where $S$ is the flux density of the specified component, $d_{\rm L}$ is the luminosity distance at the redshift $z$, and $\alpha$ is the spectral index assumed to be $\alpha\equiv\alpha_1=0$ for total VLBA and unresolved core components, and $\alpha_2=-0.7$ for the jet ($S_{\rm \nu}\sim \nu^{\alpha}$; Lister \& Homan 2005). The rest-frame ratios of $C_{\rm VLBA}=S_{\rm VLBA}/S_{\rm tot}$ (where $S_{\rm tot}$ is the flux density at 15\,GHz determined from observations of single dish antennas) and $C_{\rm un}=S_{\rm un}/S_{\rm VLBA}$ define the source compactness on arcsecond scales and sub-milliarcsecond scales respectively \citep{kovalev05}. The radio-loudness parameter is defined as the ratio of the total VLBA flux density at 15~GHz and the nuclear optical flux measured in the B band (see Eq.~\ref{eq:fb}), $R=S_{\rm VLBA}(1+z)^{-\alpha_1+\alpha_3}/F_{\rm B, \,nuclear}$, where $\alpha_3$ is the optical spectral index adopted to be $-0.5$ (see section \ref{sec:optLum}). 

We measure the intrinsic luminosity of the jet for 68 AGN (51 quasars and 12 BL Lacs) using the relation $L_{\rm int}=D_{\rm var}^{-(p-\alpha)} L$, where $p=2$ for a steady-state jet, appropriate for a core region, and $p=3$ for a discrete optically thin source \citep{lind85}, and $\alpha\equiv\alpha_1=0$ for $L_{\rm VLBA}$, $L_{\rm un}$, and $\alpha\equiv\alpha_2=-0.7$ for $L_{\rm jet}$. 

The apparent speed of the jet ($\beta_{\rm a}$ in units of the speed of the light) is measured for 118 AGN \citep[][ and MOJAVE website\footnote{http://www.physics.purdue.edu/MOJAVE/}; see Table~\ref{tab:3}]{lister09b}. We use the variability Doppler factor of the jet ($D_{\rm var}$; Hovatta et al. 2009) to derive the Lorentz factor ($\gamma=1/\sqrt{1-\beta^2}$, where $\beta$ is the speed of the jet in units of the speed of the light),
\begin{equation}
 \gamma = \frac{\beta_{\rm a}^2+D_{\rm var}^2+1}{2D_{\rm var}},
\end{equation}
and the viewing angle of the jet
\begin{equation}
 \theta = \arctan \left(\frac{2\beta_{\rm a}}{\beta_{\rm a}^2+D_{\rm var}^2-1}\right)
\end{equation}
for 63 AGN from the MOJAVE-1 sample. Note that this value represents the 49\% of the total of AGN.
We cannot estimate the exact errors of the Lorentz factor and the jet viewing angle, because it is difficult to determine the error estimates of $D_{\rm var}$ for each source. \cite{hovatta09} used a sample of 45 AGN with well-defined radio flares to determine the median standard deviation of the sample to be $\sim 27$\%, which is assumed to be an upper limit for the error estimate of $D_{\rm var}$. Adopting this value we calculate the upper limit of the error imposed on the median value of $D_{\rm var}$ for the MOJAVE-1 sample, $D_{\rm var,\,med}=12.2 \pm 3.31$. Using the later and the median apparent speed and the median error of the sample, $\beta_{\rm a,\,med}=10.24 \pm 0.81$, we determine the upper limits of the error estimates for $\gamma$ and $\theta$ to be $\sim 8$\,\% and $\sim 32$\,\%, respectively, using the formula for error propagation of two uncorrelated variables, $\beta_{\rm a,\,med}$ and $D_{\rm var,\,med}$.

To measure the power transported by the radio jet to kiloparsec scales, we used the analytical formula derived by \citet{punsly05},
\begin{equation}
 Q_{\rm j} = 5.7\times10^{44}(1+z)^{1+\alpha}Z^2S_{\rm 151}\,\,{\rm erg\,s^{-1}},
 \label{eq:qj}
\end{equation}
where $S_{151}$ is the flux density at 151\,MHz detected from the extended radio structure (see Table~\ref{tab:3}), $\alpha=1$ is the spectral index at 151\,MHz, and $Z$ is a function of the redshift. The flux densities at 151\,MHz are estimated for 135 AGN from the interpolation between flux densities measured at frequencies between 80\,MHz to 1\,GHz using data available from the CATS database - Astrophysical CATalogs support System (Trushkin et al. 2000; Verkhodanov et al. 1997, 2000, 2005, 2009). The contribution of the VLBI core to the flux density at 151\,MHz should be small or negligible for the majority of compact AGN because the core emission should be synchrotron self-absorbed and/or suffer free-free absorption at low frequencies \citep{lobanov98}. Moreover, we used flux-density measurements of single-dish telescopes which, at low radio frequencies, pick up predominantly emission from the extended kpc-scale structure. The radio luminosity at 151\,MHz ($L_{151}$) is calculated from Eq.~(\ref{eq:lum}) for the value of $\alpha=-1$.

\section{Optical nuclear luminosities of compact AGN}
\label{sec:optLum}
We estimate the optical luminosities of
compact radio sources at 5100 \AA\,($L_{5100}$) using the B band in
the standard Johnson's photometric system ($\rm B_{J}$). The absolute luminosity at 5100\AA\ is estimated with the following expression \citep[e.g.,][]{marziani03a}:

\begin{equation}
\lambda L_{5100}=
3.137\times10^{35-0.4(M_{\rm {B}}-A_{\rm B})}\,
\mathrm{erg\,s^{-1}},
\end{equation}
where $A_{\rm B}$ is the galactic extinction in the B-band taken
from the NASA Extragalactic Database and $M_{\rm B}$ is the absolute magnitude given by \cite{SG83}:
\begin{eqnarray}
M_\mathrm{B}= {\rm B_{J}} - 5
\,\mathrm{log\,}A(z)+2.5\,(1+\alpha)\,\log\,(1+z)+ \nonumber \\
5\,\log\,(h)-42.386,
\end{eqnarray}
where $h = H_{0}/100$\,km\,s$^{-1}$\,Mpc$^{-1}$ = 0.7, $\alpha\equiv\alpha_3 = -0.5$ is the
spectral index  found for radio-loud objects (S$_{\nu}\propto\nu^{\alpha}$; see Brotherton et al. 2001 Table 2), $z$ is the redshift and
$A(z)$ is the bolometric luminosity distance for the flat cosmology model,
\begin{equation}
A(z)=(1+z)\left[\eta(1,\Omega_{\rm m})-\eta\left(\frac{1}{1+z},\Omega_{\rm m}\right)\right],
\end{equation}
where $\Omega_{\rm m}$=0.3 and $\eta$ is a function of $z$ and $\Omega_{\rm m}$ (see Pen 1999).

\begin{figure}
  \begin{center}
   \includegraphics[width=0.45\textwidth]{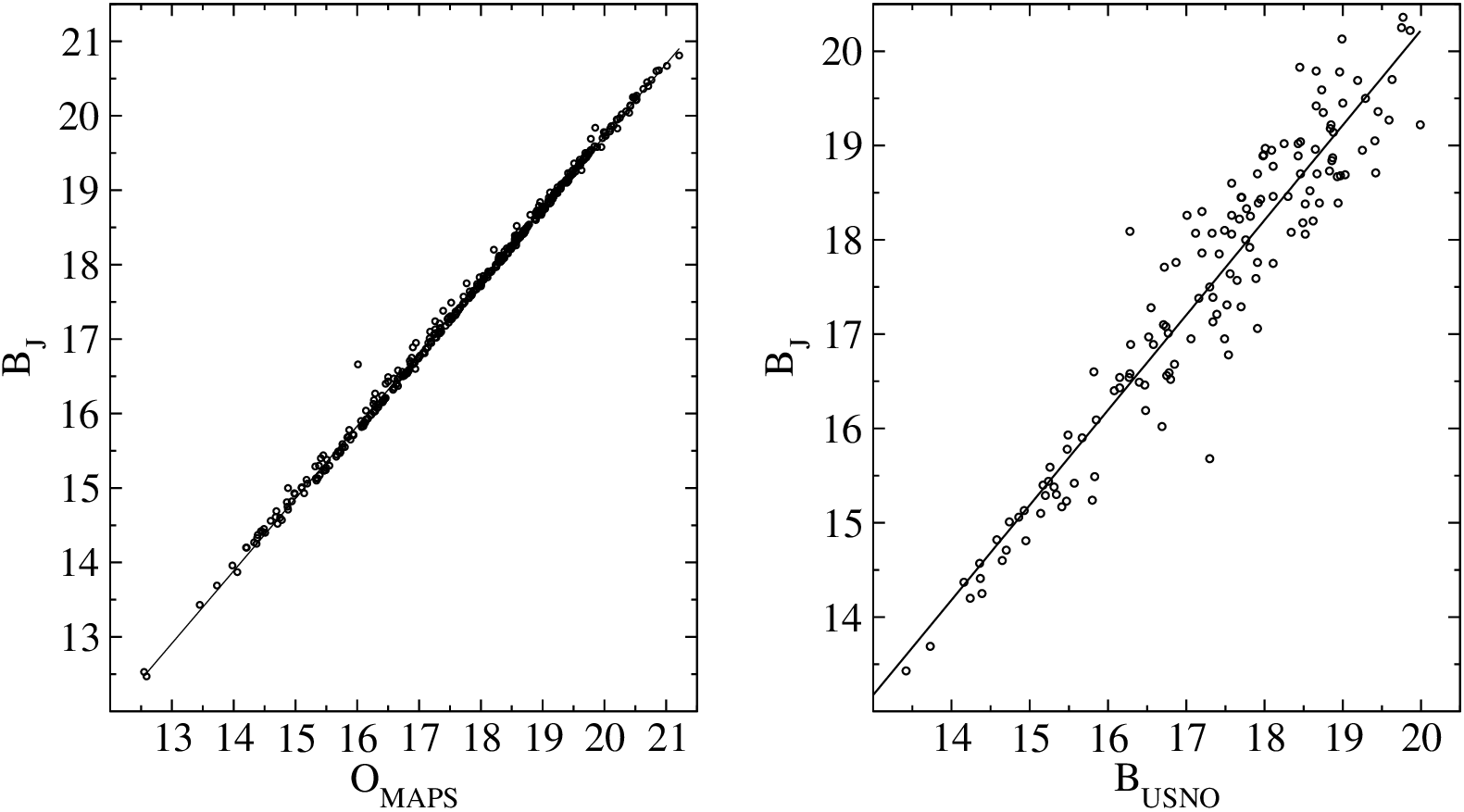}
  \end{center}
  \caption{\small{\emph{Left panel:} Linear transformation between O MAPS and $\rm B_{J}$. \emph{Right panel:} Linear transformation between B USNO and $\rm B_{J}$.}}
  \label{fig:OB.BB}
\end{figure}

We have searched for the apparent magnitudes for 250 compact AGN in two
catalogs: the Minnesota Automated Plate Scanner Catalog of the
Palomar Observatory Sky Survey (POSS I), hereafter the MAPS Catalog
\citep{cabanela03} and the US Naval Observatory
USNO--B catalog \citep{monet03}. The apparent
magnitudes in the MAPS catalog are given as photographic O
magnitudes, while the USNO catalog measures B magnitudes. To convert
the O magnitude from MAPS and the B USNO to the standard $\rm B_J$ magnitude
in Johnson's system we used the linear correlations that we found
using the BVR photoelectric photometry for 426 stars in 49 Palomar
Sky Survey fields \citep{humphreys91}. The following expressions are used to transform the apparent magnitude to the $\rm B_J$ magnitude:

\begin{figure*}
  \begin{center}
\includegraphics[width=0.65\textwidth]{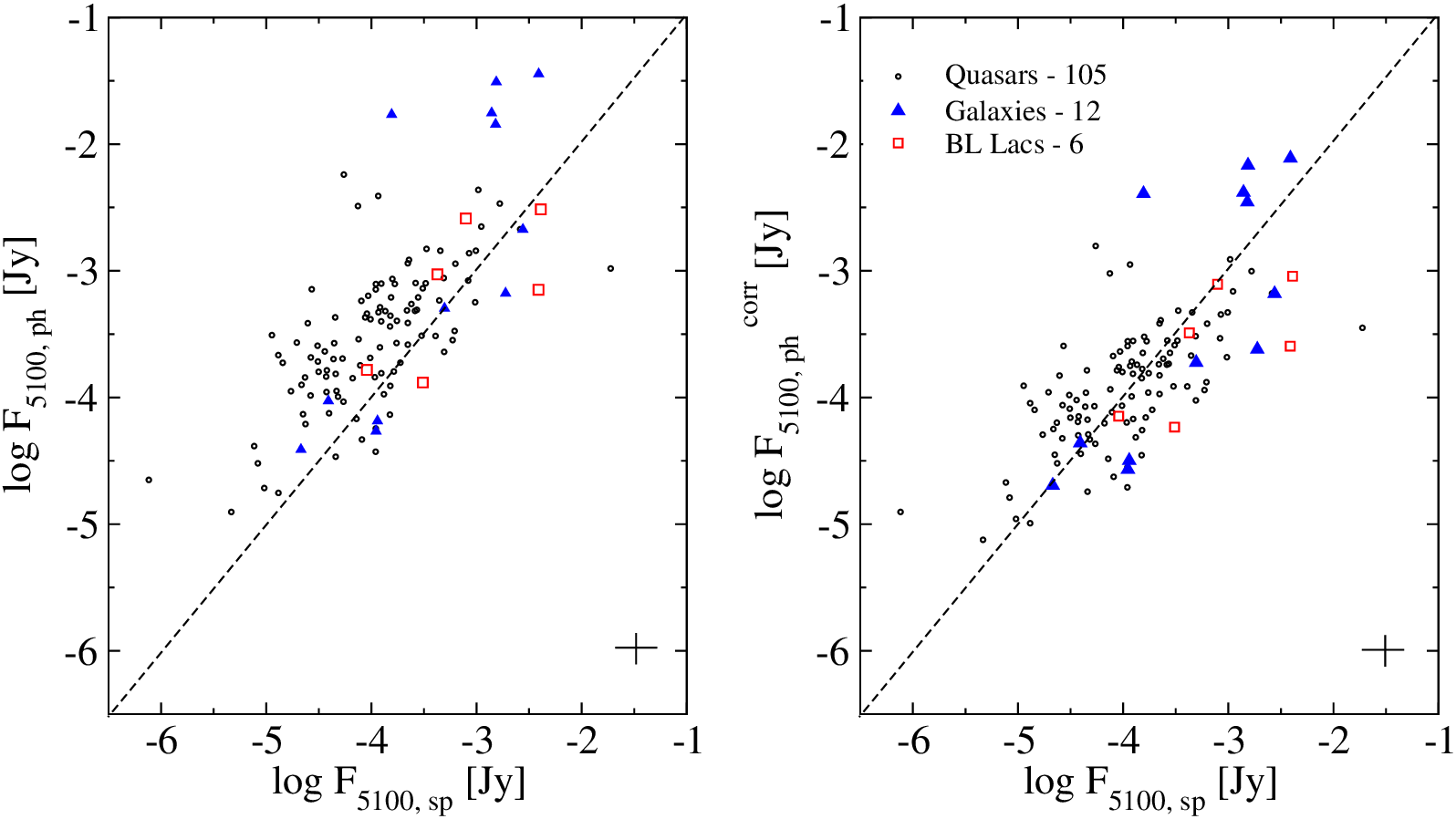}
  \end{center}
  \caption{\small{\textit{Left panel}: Photometric fluxes against spectroscopic fluxes scaled to 5100\,\AA\ for 123 AGN from the full sample. \textit{Right panel}: Corrected photometric fluxes against spectroscopic fluxes at 5100\,\AA\, for the same sample. The dotted line represents the identity. Averaged 1$\sigma$ error bars are presented (errors of individual sources are given in the Table~\ref{tab:3}).  }}
\label{fig:fph-fsp_2}
\end{figure*}

\begin{equation}
\mathrm{B_{J}}=(0.264\,\pm\,0.038) + (0.973\,\pm\,0.002)\mathrm{O_{MAPS}},
\end{equation}
and,
\begin{equation}
\mathrm{B_{J}}=(0.08\,\pm\,0.46) + (1.01\,\pm\,0.03) \mathrm{B_{USNO}}.
\end{equation}
From Fig.~\ref{fig:OB.BB} we can see that the O MAPS magnitude presents less dispersion than B USNO magnitude with respect to the $\rm B_J$.  Out of 250 sources, we
identified 163 AGN in the MAPS catalog, the other 79 remaining sources were found in the USNO catalog, and eight objects have no optical counterparts. Out of 135 sources from the MOJAVE-1 sample, nine sources have no optical counterparts: two quasars, four BL Lacs, and three unclassified objects (see Table~\ref{tab:3}). 

For 123 AGN from the full sample, we have measurements of monochromatic spectral fluxes at 5100\,\AA, 3000\,\AA, and 1350\,\AA . The spectroscopic observations were carried out at 2.1-m class telescopes OAN-SPM\footnote{Observatorio Astron\'omico Nacional at San Pedro M\'artir, Baja California, M\'exico}, GHAO\footnote{Observatorio Astrof\'isico Guillermo Haro at Cananea, Sonora,
M\'exico} and  complemented with  intermediate and high-resolution spectra found in SDSS/HST--FOS archives \citep{marziani03b,lawrence96}. The spectral atlas and the details of the observations will be published in a forthcoming paper (Torrealba et al., in preparation). 

We scaled the 3000\,\AA\, and 1350\,\AA\, fluxes to 5100\,\AA\, $F_{\rm 5100, sp}$ assuming the spectral index $\alpha\,=-\,0.5$. Contribution of optical nuclear emission is dominant for quasars and BL Lacs, while for radio galaxies the optical emission of the stellar component can be significant.
Figure~\ref{fig:fph-fsp_2} (\textit{left panel}) shows that the photometric fluxes ($F_{\rm 5100}$) are on average higher than the spectral fluxes ($F_{\rm 5100, sp}$), because the contribution of the host galaxy is larger in photometric fluxes. To correct the photometric fluxes for a stellar emission, we used the relation between $F_{\rm 5100}$ and $F_{\rm 5100, sp}$ for 123 AGN, excluding the radio galaxies and eight highly variable AGN:

\begin{equation}
\label{eq:fb}
 F_{\rm B, \,nuclear} \equiv F_{\rm B, \,corr} = 0.14\,(F_{\rm B})^{0.871\pm0.072}.
\end{equation}

Corrected photometric fluxes, measured and estimated spectral fluxes at 5100\,\AA\, are presented in Fig.~\ref{fig:fph-fsp_2} (\textit{right panel}). Outliers from the equality line are radio galaxies and few quasars. The corrected fluxes of the former are contaminated by the host galaxy emission, while a strong variability of some quasars causes large deviations expected from the equality line.

\begin{figure}
  \begin{center}
  \includegraphics[width=0.45\textwidth]{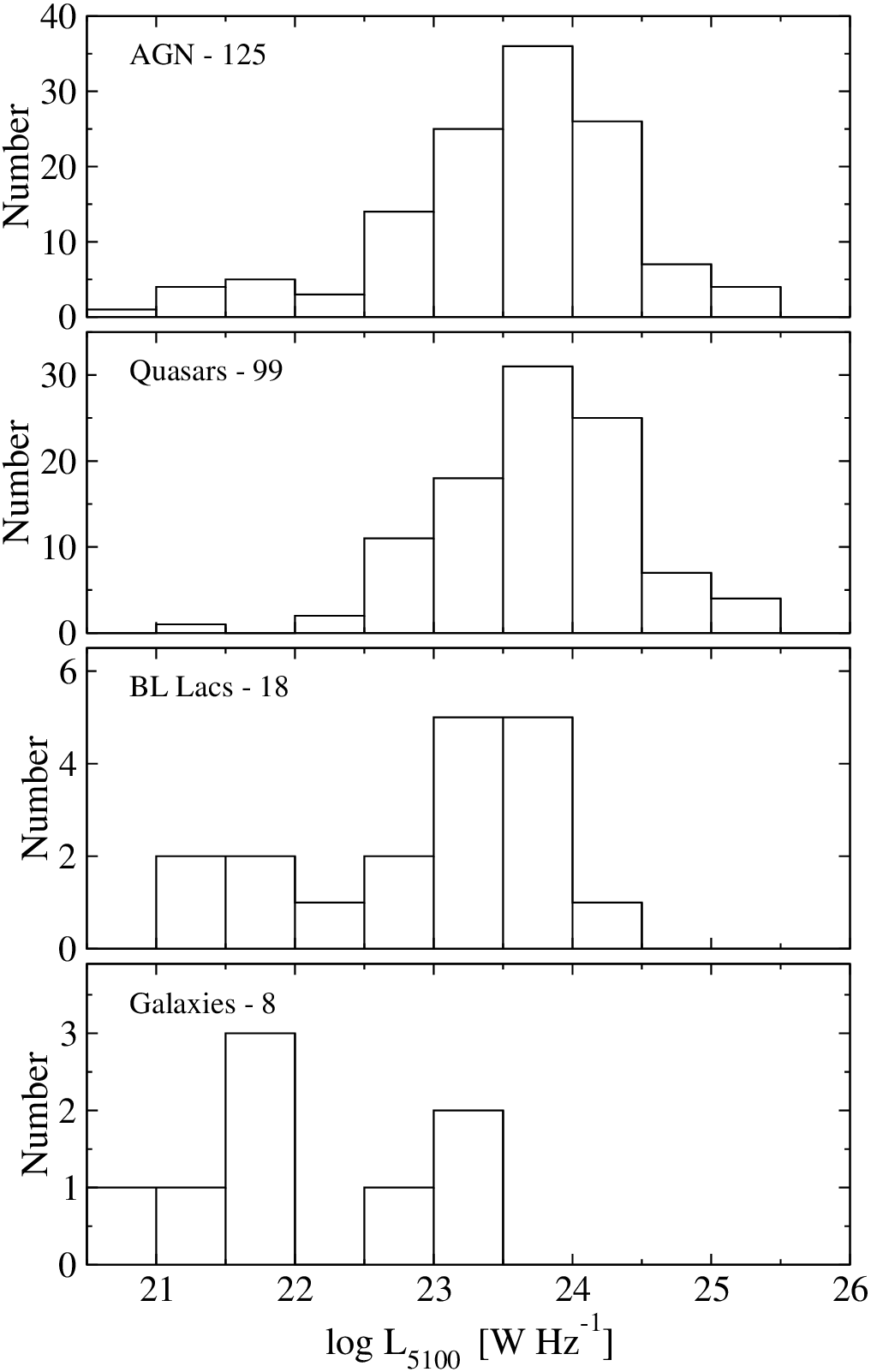}
  \end{center}
  \caption{\small{Distribution of corrected photometric optical luminosities at 5100\,\AA\ for the sample of 125 AGN for MOJAVE-1 sample (top panel), 99 quasars, 18 BL Lacs and 8 radio galaxies}.}
\label{fig:hist_lumo}
\end{figure}

The corrected optical fluxes and corrected luminosities (or nuclear optical luminosities, hereafter, $L_{5100}$) are presented in Table~\ref{tab:3}.
The distribution of nuclear optical luminosities of 123 AGN from the MOJAVE-1 sample ranges between $10^{20}$\,W\,Hz$^{-1}$ and $10^{25.5}$\,W\,Hz$^{-1}$ with the maximum at $10^{24}$\,W\,Hz$^{-1}$  (Fig.~\ref{fig:hist_lumo}). Optical luminosities of quasars are higher than those of BL Lacs, and the luminosities of the latter are higher than those for radio galaxies. While the optical luminosities are weaker than the total radio luminosities by a three orders of magnitudes, the shapes of distributions of $L_{5100}$ for quasars, BL Lacs and galaxies are similar to the distributions of total radio luminosities of the compact jets at 15\,GHz \citep{lister05,kovalev05}.

\section{Correlations between radio properties of the jet}
\label{sec:rCorr}
Correlations are considered significant for the samples of all 135 sources and 95 quasars if the confidence level $P>98\%$, while for the sample of 18 BL Lacs the correlations with a confidence level high than 95\% are considered. 
Partial Kendall's tau correlation \citep{akritas96} showed that the VLBA luminosities of MOJAVE-1 AGN and the luminosities of their unresolved core and the jet at 15\,GHz are correlated at the high confidence level, $>99.99\%$. These correlations were expected for all types of AGN such as quasars, BL Lacs and radio galaxies, because the emission from both unresolved core and jet are expected to be Doppler-boosted.

\begin{figure}
  \label{fig:lumr-theta}
  \begin{center}
  \includegraphics[width=0.45\textwidth]{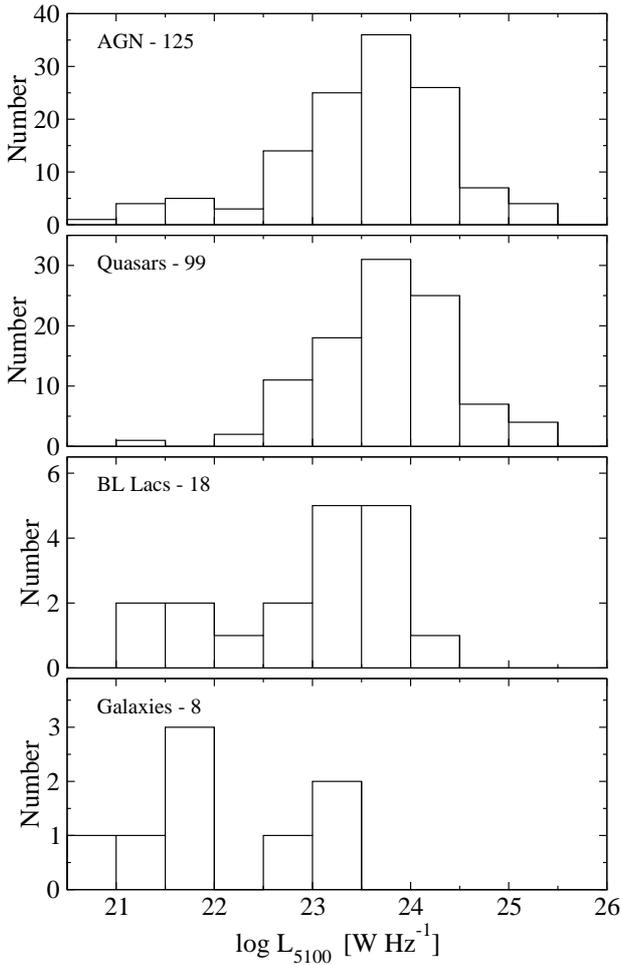}
  \end{center}
  \caption{\small{Viewing angle of the jet against radio luminosity of the VLBA jet at 15\,GHz for 48 quasars, 12 BL Lacs, and 5 radio galaxies.}}
\end{figure}

In the relativistic brightening model the observed radio emission of the jet is boosted by relativistic Doppler effect, which depends on the viewing angle ($\theta$) and the Lorentz factor ($\gamma$) of the jet. Strong negative correlation  between $L_{\rm VLBA}$, $L_{\rm un}$, $L_{\rm jet}$ and $\theta$ for 48 quasars (Table~\ref{tab:1}) and for the sample of 63 AGN (Fig.~\ref{fig:lumr-theta}) confirms that the radio emission from all AGN types suffers from relativistic beaming. The missing population of compact (core only) jets without viewing angle ($\theta$) measurements are those aligned to the line-of-sight and should have populated the right bottom quadrant in Fig.~4. The ordinary linear-squares method is used to fit this relation for 62 AGN, excluding three outliers (two quasars and one BL Lac),
\begin{equation}
  \log \theta = (7.92\pm0.78) + (-0.26\pm0.03)\log L_{\rm VLBA}.
\end{equation}
This relation can be used as an independent method for estimating the jet viewing angle from radio luminosity of the jet at 15\,GHz.

No correlation is found between jet luminosities and jet speed separately for quasars, BL Lacs and radio galaxies. The average Lorentz factors of quasars and BL Lacs ($17.1\pm10.9$ and $8.2\pm5.5$) are larger than those for radio galaxies ($3.6\pm2.9$). This produces a positive correlation ($95\,\%$) in the $L_{\rm VLBA}-\gamma$ plane, which appears to be stronger between $L_{\rm jet}$ and $\gamma$ ($98.4\,\%$; Table~\ref{tab:2}).


\begin{figure}
  \label{fig:hist_Qjet}
  \begin{center}
\includegraphics[width=0.45\textwidth]{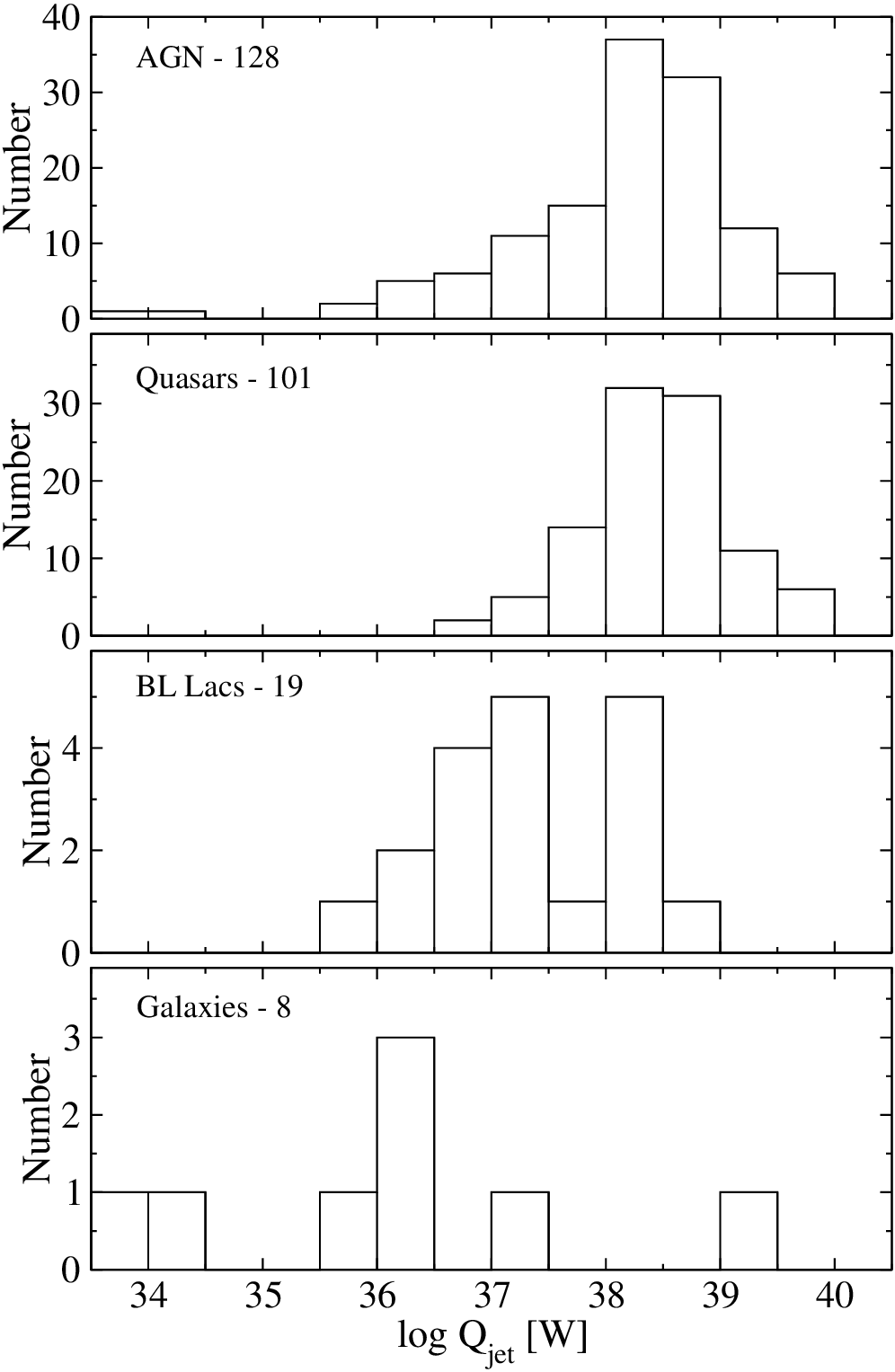}  
  \end{center}
  \caption{\small{Distribution of the jet kinetic power for 128 AGN (top panel): 101 quasars, 19 BL Lacs, and 8 radio galaxies.}}
\end{figure}

\begin{figure}
  \label{fig:Qjet}
  \begin{center}
\includegraphics[width=0.45\textwidth]{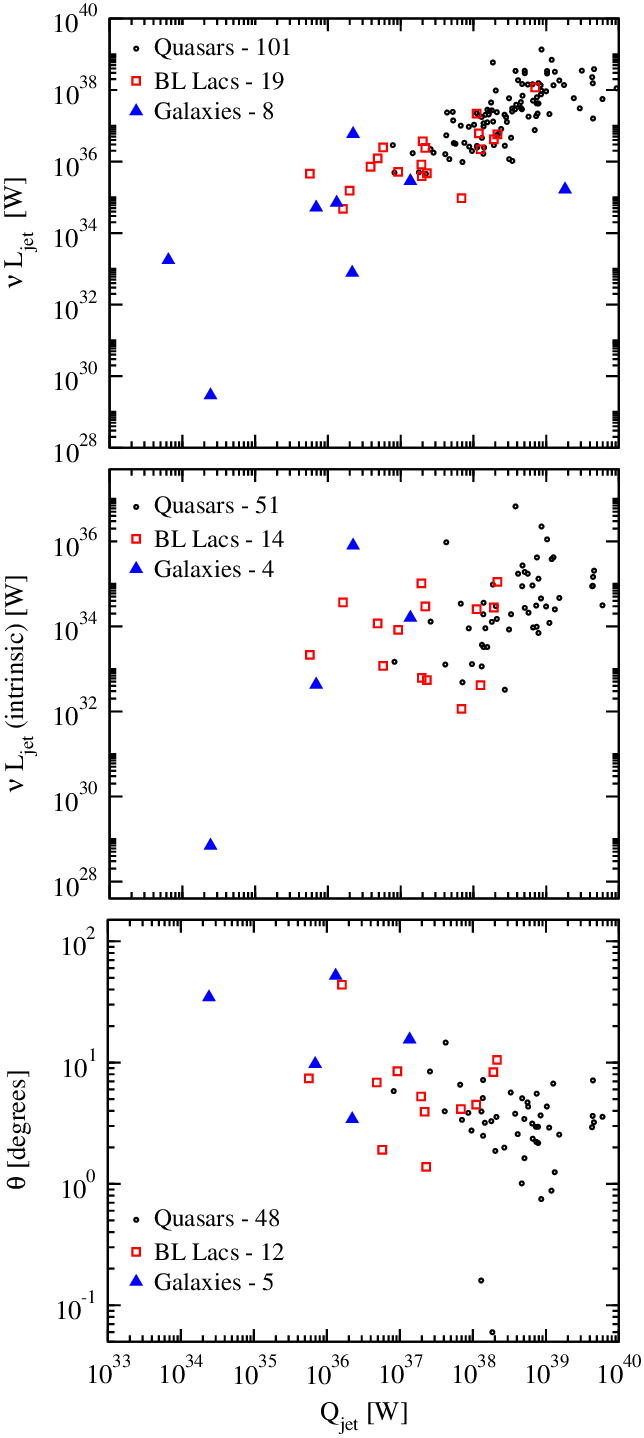} 
  \end{center}
  \caption{\small{Jet kinetic power versus apparent luminosity (top panel) and intrinsic luminosity (middle panel) of the jet at 15\,GHz, and viewing angle of the jet (bottom panel). Note that the upper limit of errors associated with the jet viewing angle is $\sim 30$\%. }}
\end{figure}

The distribution of the jet kinetic power ($Q_{\rm j}$ estimated from Eq.~(\ref{eq:qj})) is significantly different for quasars, radio galaxies, and BL Lacs (Fig.~\ref{fig:hist_Qjet}). We should not expect a difference between quasars and radio galaxies as they are according to the unification scheme the same objects oriented differently to the line-of-sight. In the MOJAVE sample, the kinetic power of the jets are found to be higher in quasars than in radio galaxies. This can be understood as a bias resulting from the flux limit of the sample: only nearby radio galaxies ($z<0.2$) are included in the MOJAVE sample because they are fainter at 15\,GHz, while quasars are more luminous and can be detected up to $z\approx2.5$.

The negative correlation, at 99.7\,\% confidence level found for all types reflects the fact that BL Lacs and radio galaxies have on the average smaller kinetic powers and larger jet viewing angles than those for the quasars. 

We find a strong positive correlation between the kinetic power of the jet and the apparent luminosity of the jet for 101 quasars and 19 BL Lacs (Fig.~\ref{fig:Qjet}). At first glance it seems to be a surprising result because $Q_{\rm j}$ is thought to be an orientation-independent parameter, while $L_{\rm VLBA}$ or $L_{\rm jet}$ ($L_{\rm VLBA} \propto L_{\rm jet}$) depends on the jet viewing angle (see Table~\ref{tab:1}). As was discussed in Sect.\ref{sec:rsample}, more luminous jets are found at high redshifts due to their rarity.  
The $L_{\rm VLBA}$ depends on the Doppler factor of the jet (or its jet viewing angle and Lorentz factor) and the intrinsic radio luminosity,  $L_{\rm jet}=D_{\rm var}^{2-\alpha}L_{\rm jet,\,int}$ (for a steady-state jet). While progressively smaller jet viewing angles are detected at high redshifts (see Table~\ref{tab:1}), the average Doppler factor does not change much (Lister \& Marscher 1997; Fig.~13 in Hovatta et al. 2009). This indicates that the high apparent luminosities detected at high redshift are mostly due to high intrinsic luminosities of the jets. Hence, we should expect that $L_{\rm jet, \,int}$ correlates with $Q_{\rm j}$. For a sample of 51 quasars, the intrinsic luminosity of the unresolved core and the jet are found to be correlated with the kinetic power of the jet at $\ga99$\% confidence level. The significance is lower than that found in the $Q_{\rm j}-L_{\rm jet}$ relation plane (see Table~\ref{tab:1}) mainly due to about two times smaller sample of AGN with known Doppler factor values. This positive correlation may reflect that intrinsically much brighter jets on parsec-scales pump more electrons into the kiloparsec-scale radio lobes, resulting in a more luminous extended structure. A larger number of low energy electrons accumulate in the lobes (because the lifetime of the low energy electrons is much longer) and emits a low-frequency synchrotron radio emission on scales of hundreds of kiloparsecs.  We suggest that a real positive correlation between the intrinsic luminosity of the parsec-scale jet and the kinetic jet power (or luminosity of the extended radio structure) is reflected in the  $Q_{\rm j}-L_{\rm jet}$ relation plane for a given distribution of Lorentz factors of the jet.

\begin{figure}
  \label{fig:Qjet1}
  \begin{center}
  \includegraphics[width=0.45\textwidth]{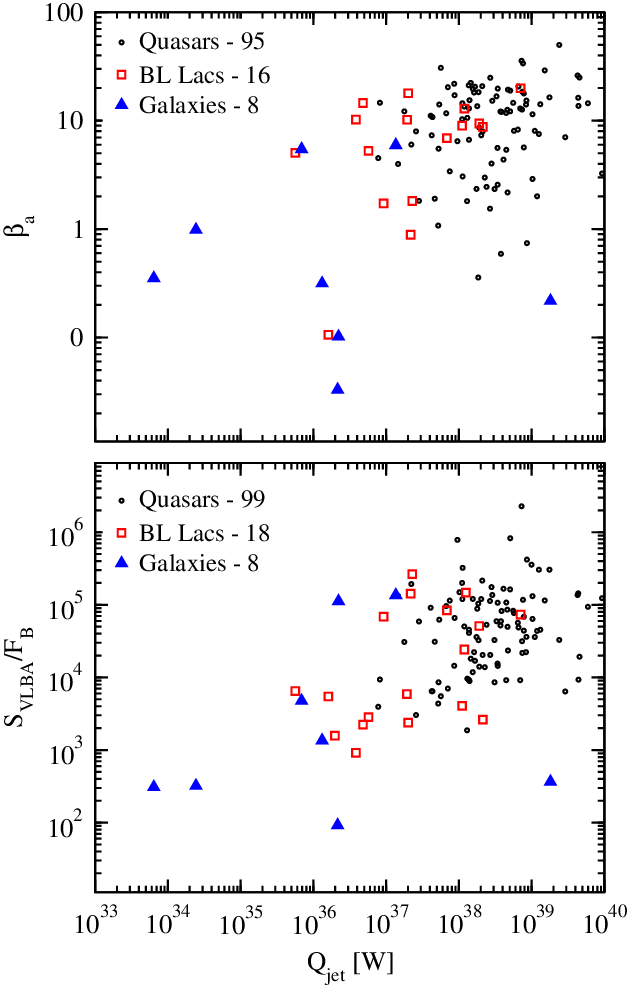}
  \end{center}
  \caption{\small{Jet kinetic power versus apparent speed and radio loudness of the VLBA jet.}}
\end{figure}

We further investigate correlations between $Q_{\rm j}$ and the intrinsic parameters of the jet ($\theta$ and $\gamma$) measured for $\sim 50$ AGN. The measured kinetic power should be independent of the jet viewing angle. This is consistent with the lack of correlation found in the samples of 48 quasars and 12 BL Lacs (Table~\ref{tab:1}). This correlation becomes significant for 46 quasars ($\tau=-0.22$, $P=0.03$) if we exclude two quasars with very small viewing angles ($<0.2$ degrees). The lack of large viewing angles in quasars with low kinetic powers (Fig.~\ref{fig:Qjet1}; bottom panel) can be explained as a selection imposed by the flux limit: (a) quasars with large kinetic power can be found at high redshifts due to their rarity, and (b) at large distances only the most beamed sources (small viewing angles and intrinsically luminous) can be included in the sample. These two effects can produce the negative correlation observed in the $Q_{\rm j}-\theta$ relation plane.
Another possibility to explain this relation is that the radio emission at 151\,MHz is variable and relativistically beamed. In the minority of the low-frequency variable sources (LFV), the variability is correlated with that at high frequencies, with events started at high frequencies and propagated to low frequencies with decreasing amplitude. This is a signature of intrinsic variability caused by expansion of the emitting region, which has been confirmed from a comparison between flux density variability at 408\,MHz and structural changes observed with VLBI observations at 1.67\,GHz in several LFV sources \citep{bondi96}. The majority of sources selected on the basis of LFV, however, showed no correlation between variability at low and high frequencies, which has been interpreted in terms of extrinsic variability due to refractive interstellar scintillation \citep[e.g.,][ and references therein]{bondi94}. Since the MOJAVE sources are selected from high-frequency radio surveys on the basis of their beamed emission, it is most likely that the low frequency variability is a result of both beaming and refractive interstellar scintillation, with the former being dominant. Coordinated monitoring of MOJAVE blazars across the wide range of frequencies is needed to explore this issue. To fully understand all selection effects present in the flux-limited MOJAVE-1 sample and their role in shaping the observed correlations, it would be helpful to simulate the selection effects and the correlations between radio parameters.

The Lorentz factors also do not show a significant correlation with $Q_{\rm j}$ for quasars and BL Lacs when considered separately (the significant positive correlation present for the joint sample is a result of jet power and jet speed being on average larger in BL Lacs than in quasars). 

The relationship between intrinsic jet luminosity and extended lobe
power is currently not well-constrained (Lister \& Homan 2005). We report a positive correlation found for 51 quasars (98\,\%) between the intrinsic luminosity of the parsec-scale jet and the luminosity of the extended radio structure suggesting that intrinsically bright parsec-scale jets produce luminous radio lobes on scales of hundreds of kiloparsecs.

Because of the tight correlation between $L_{\rm 151}$ and $Q_{\rm j}$, the jet kinetic power  is also correlated with $L_{\rm jet,\,int}$ (99\,\%) and $L_{\rm un,\,int}$ (99\,\%).
Positive correlations are found between $Q_{\rm j}$ and $S_{\rm VLBA}/F_{\rm B}$, the radio loudness of the jet. This correlation is significant for 99 quasars ($99.1\,\%$) and marginally significant for 18 BL Lacs (94\,\%).

Another significant negative correlation (99.9\,\%) exists for quasars between the kinetic power of the jet and the compactness of the core on sub-milliarcseconds scales, while it is positive for BL Lacs with a confidence level of 95.4\,\%. Larger samples of BL Lacs are needed to check the later correlation. There is a strong positive correlation at 99.9\,\% level between $Q_{\rm j}$ and $\beta_{\rm a}$ for 119 AGN (95 quasars), meaning that the kinetic power is correlated with the speed of the jet, $\gamma$. We checked this for 48 quasars: the correlation coefficient is positive but the correlation is not significant. A larger sample of quasars with measured Doppler factors is needed to detect a significance between weakly correlated parameters.

The kinetic power of the jet depends on the Lorentz factor of the jet and the mass outflow rate, $Q_{\rm j}=\gamma \dot{M}_{\rm out}c^2$, where $c$ is the speed of the light. This opens a possibility to measure the mass outflow rate using the independent measurements of $Q_{\rm j}$ and $\gamma$ and to use the $\dot{M}_{\rm out}$ as a measure of the accretion rate in superluminal AGN. We
will pursue this study in a separate paper.

\section{Optical continuum luminosity and properties of the jet for MOJAVE-1 sources}
\label{sec:roCorr}
Compact AGN are strongly variable at 15\,GHz
and in the optical band \citep{kovalev05,wold07}. In our case of non-simultaneous observations, variability can blur possible correlations between radio and optical luminosities.

The total radio luminosity of the VLBA jets at 15\,GHz against the optical continuum luminosity at 5100\,\AA\ for quasars, BL Lacs, and radio galaxies are shown in Fig.~8.

The partial Kendall's tau correlation analysis is used to investigate correlations between radio and optical luminosities. We find a significant
positive correlation ($>99.99\,\%$) between the nuclear optical luminosity of compact AGN and the radio luminosities of the total VLBA emission, the unresolved core, and the jet emission  for 99  quasars (Fig.~8; Table~\ref{tab:2}), while for 18 BL Lacs, the optical luminosity correlates positively only with the jet luminosity at the confidence level of $\sim99.9\,\%$. 
According to the unification scheme for radio-loud AGN, the jet viewing angles are smaller in BL Lacs than in quasars and it is larger in radio galaxies, but this picture is not consistent with the average angles, $\theta_{\rm j}$, estimated for quasars, BL Lacs and radio galaxies from the MOJAVE sample, $(3.7\pm 2.5)$ deg, $(8.9\pm 11.3)$ deg and $(23.1\pm 19.9)$ deg, respectively. This is most likely a selection effect - BL Lacs viewed at smaller angles are more compact, and for these we cannot measure jet viewing angles. This introduces a bias for the sample of BL Lacs towards larger viewing angles. \cite{ghisellini05} suggested that there is a stratification of velocities in the jet of BL Lacs. The emission of the fast-spine layer is strongly boosted into a narrow cone and only the outer lower-speed layers are detected, which are not representative of the characteristic speed of the jet, and, hence, of the Doppler factor. This effect, together with possible misclassification of some BL Lacs and quasars \citep[see][]{hovatta09} may lead to a bias towards large viewing angles for BL Lacs.  

\begin{figure}
  \label{fig:lr-lo}
  \begin{center}
 \includegraphics[width=0.45\textwidth]{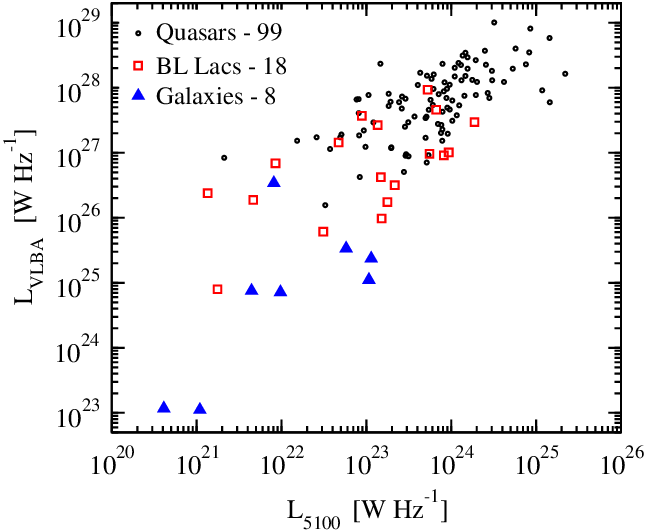}  
  \end{center}
  \caption{\small{Total VLBA luminosity at 15\,GHz versus optical nuclear luminosity at 5100\,\AA\ for 125 compact AGN.}}
\end{figure}

\begin{figure}
  \begin{center}
    \includegraphics[angle=-90,width=0.48\textwidth]{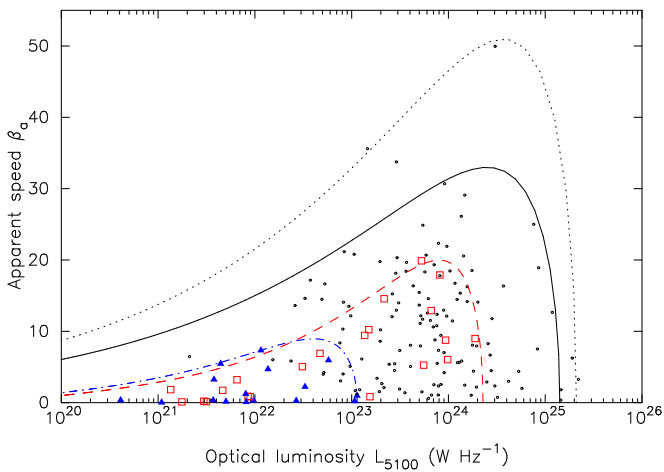}
  \end{center}
  \caption{\small{Apparent speed of the jet versus optical nuclear luminosity at 5100\,\AA\ for 164 AGN from the MOJAVE sample. The sample of AGN consists of 127 quasars (circles), 21 BL Lacs (open squares), and 16 radio galaxies (filled triangles). The aspect curve is given for quasars (dotted and full lines which cover all and 95\% of quasars, respectively), BL Lacs (dashed line), and radio galaxies (dot-dashed line). 
}}
    \label{fig:aspeed-olum}
\end{figure}

Radio-optical correlations for quasars and BL Lacs can be understood if both radio and optical synchrotron emission originate in the relativistic jet and are Doppler-boosted. The dispersion of the radio-optical correlation can be caused by non-simultaneous observations, distributions of intrinsic luminosities, and Doppler factors. The larger dispersion for BL Lacs could be a result of stronger variability in the radio and optical bands as well as a wider range of intrinsic luminosities, while for radio galaxies the dimming of optical continuum emission by an obscuring dusty torus can vary significantly.

\cite{arshakian10} reported a correlation between the ejected jet components
and the optical continuum flares in the radio-loud galaxy 3C\,390.3. This correlation was confirmed for another radio galaxy, 3C\,120 \citep{tavares10}, suggesting that this link is common for all radio-loud galaxies and radio-loud quasars. The link between optical continuum variability and kinematics
of the parsec-scale jet is interpreted in terms of optical
flares generated at subparsec-scales in the innermost part of a relativistic jet \citep{arshakian10,tavares10} rather than in the accretion disk. This suggests that the bulk of optical continuum emission is non-thermal and Doppler-boosted. It is therefore likely that the optical nuclear luminosity is intrinsically correlated with the radio luminosity of the unresolved core of radio galaxies and quasars rather than with their jet or accretion luminosities. For the BL Lacs, the significant correlation found in the $L_{5100}-L_{\rm jet}$ relation plane suggests that the optical emission is likely to be generated in the continuos jet. 

Another evidence supporting the synchrotron nature of the optical emission generated in the relativistic jet is coming from the apparent speed--optical luminosity diagram (Fig.~\ref{fig:aspeed-olum}). There is a lack of blazars with low optical luminosities and high apparent speeds, similar to the absence of sources in the relation between apparent transverse speed and apparent radio luminosity of the MOJAVE sources at 15\,GHz \citep{cohen07}. This is not a selection effect due to the observing limit in the MOJAVE-1 sample and it is interpreted in terms of relativistic beaming of radio jets. The aspect curve, the relation between $L_{\rm VLBA}$ and $\beta_{\rm a}$  for a fixed intrinsic radio luminosity and Lorentz factor of the jet, derived from the relativistic beaming theory were shown to be a good envelope for the quasars in the 
$L_{\rm VLBA}-\beta_{\rm a}$ diagram \citep{cohen07}, suggesting that the relativistic beaming model is realistic for the MOJAVE blazars. 

\begin{figure}
  \begin{center}
  \includegraphics[width=0.45\textwidth]{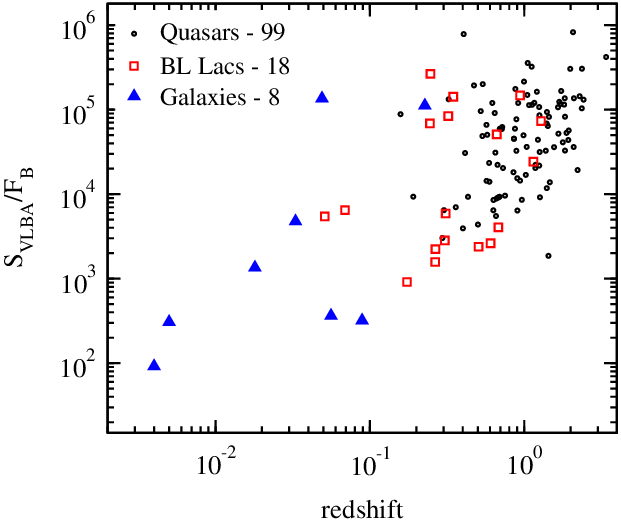}
  \end{center}
  \caption{\small{Radio-loudness versus redshift for 125 compact AGN.}}
  \label{fig:r-z}
\end{figure}

The good fit of the aspect curve in the $L_{5100}-\beta_{\rm a}$ diagram is shown for the whole population of quasars and 95\,\% of the population (dotted and full lines in Fig.~\ref{fig:aspeed-olum}). The good fit of an aspect curve for all quasars suggests the peak values of the intrinsic optical luminosity $L_{\rm 5100,\,int}=2\times10^{20}$~W Hz$^{-1}$, $\gamma=52$, and $p=3$ in the population (Fig.~\ref{fig:aspeed-olum}; dotted line). However, for the majority of quasars (95\,\%), a better fit can be achieved with the peak values of $L_{\rm 5100,\,int}=4\times10^{20}$~W Hz$^{-1}$, $\gamma=33$, and $p=3$. The populations of BL Lacs and radio galaxies are also nicely enveloped under the aspect curve with parameters $L_{\rm 5100,\,int}=9\times10^{21}$~W Hz$^{-1}$, $\gamma=20$, and $p=2$ (dashed line) and $L_{\rm 5100,\,int}=1.5\times10^{21}$~W Hz$^{-1}$, $\gamma=9$, and $p=2$ (dot-dashed line). This is a strong evidence that the optical nuclear emission of quasars, BL Lacs, and radio galaxies from the MOJAVE sample is generated by relativistic particles in the jet and is Doppler boosted. The quasars data are best fitted with $p=3$, indicating that the source of optical emission is likely to be a discrete and optically thin.

We find that the radio-loudness increases with increasing redshift for quasars and radio galaxies (Fig.~\ref{fig:r-z}). The significance of correlation is $99.6\,\%$ for 95 quasars (Table~\ref{tab:1}) and is mainly due to the lack of low radio-loud quasars at high redshifts. In the MOJAVE-1 sample, the probability to find the intrinsically luminous jets and jets with larger Lorentz factors is high at high redshifts, while at redshifts less than $\simeq 0.6$ there is a deficit of sources with high Doppler factors and small viewing angles, because they are rare in the population and the volume element is small at low-$z$ \citep{lister09b}. The correlation in the $R-z$ relation plane is evident at $z\ga0.6$ and can be interpreted as a tendency for strong AGN to have a high radio-loudness. In high-redshift superluminal quasars, the beamed optical emission from the jet dominates over the optical emission from the accretion disk. In this case, we can interpret the $R-z$ correlation as the radio-strong quasars to have lower Doppler factors in the optical regime compared to the ones in radio. 
Note that in about ten low-redshift quasars ($z\la 0.6$) with mostly low Doppler factors, the optical emission from the disk can be comparable/dominant to that of the jet.



Correlations presented in Table~\ref{tab:2} between photometric optical luminosity and radio luminosity of the compact jet, kinetic jet power, and radio-loudness hold also for optical spectral luminosities measured for 61 AGN.

\section{Summary}
\label{sec:conc}
We used the statistically complete sample of 135 compact radio sources provided by the MOJAVE program to investigate correlations between the properties of the pc-scale jets at 15\,GHz, the extended radio emission at 151\,MHz, and the optical nuclear emission at 5100\,\AA . The main results are summarized as follows: 
\begin{enumerate}

  \item We determine the optical nuclei fluxes at 5100\,\AA\ for 135 MOJAVE-1 AGN with their photometric optical fluxes available in the MAPS and USNO-B catalogs. There is a significant positive correlation for 99 quasars and 18 BL Lacs between optical nuclear luminosities and jet luminosities at 15\,GHz originated in the jet at milliarcseconds scales. For quasars, correlations hold also between optical nuclear luminosity and luminosities of the unresolved core (at sub-milliarcseconds scales) and total radio (VLBA) of the jet, suggesting that the optical emission is non-thermal and originates in the innermost part of the jet at sub-parsec scales, while in the BL Lacs it is generated in the parsec-scale jet. \\

The generation of the relativistically beamed optical emission in the MOJAVE blazars is evident from the apparent speed -- optical nuclear luminosity relation plane. In this diagram, the data of quasars, BL Lacs and radio galaxies are fitted by an aspect curve derived from the relativistic beaming theory, suggesting that optical emission from these sources is relativistically Doppler-boosted. We estimate the peak values of the intrinsic optical luminosity and the Lorentz factor of the jet for each population of blazars: $L_{\rm 5100,\,int}=2\times10^{20}$~W Hz$^{-1}$, $\gamma=52$, and $p=3$ for quasars, $L_{\rm 5100,\,int}=9\times10^{21}$~W Hz$^{-1}$, $\gamma=20$, and $p=2$ for BL Lacs, and $L_{\rm 5100,\,int}=1.5\times10^{21}$~W Hz$^{-1}$, $\gamma=9$, and $p=2$ for radio galaxies. About 95\,\% of quasars have the peak values of $L_{\rm 5100,\,int}=4\times10^{20}$~W Hz$^{-1}$, $\gamma=33$, and $p=3$. The boosting factor $p=3$ favors the source of optical continuum emission for quasars to be discrete and optically thin.  \\

  \item The kinetic power of 135 compact jets, determined from the flux density measured at 151\,MHz, is on  average higher in quasars than in BL Lacs, and it is lower in radio galaxies. We find a strong positive correlation between the intrinsic kinetic power of the jet and the apparent luminosities of the total and unresolved core emission of the jet at 15\,GHz. This correlation can be a result of the correlation between kinetic power and intrinsic luminosity of the pc-scale jet, which reflects that intrinsically luminous compact jets deliver more relativistic electrons to kiloparsec-scales. In this way they accumulate more low-energy electrons in the radio lobes, which results in powerful extended radio lobes radiating at low radio frequencies. 

Another possibility to interpret the correlation found in the $Q_{\rm j}-L_{\rm jet}$ relation plane is that the low-frequency emission at 151\,MHz is relativistically beamed and its variability is correlated with beamed radio emission at 15\,GHz. Correlated variability at low and high frequencies was found in few superluminal sources. It is most likely that both effects, the beaming at low frequencies and the intrinsic luminosity of the jet at 15\,GHz in beamed AGN, contribute to the relation between kinetic power and apparent luminosity of the jet. Monitoring of the MOJAVE AGN in a wide range of frequencies and over a period of few years is needed to distinguish and estimate the contribution of each effect.

The kinetic power of quasars positively correlates with the radio-loudness, i.e., the ratio of the radio flux density at 15 GHz to the optical flux at 5100\AA\,, and negatively with the compactness of the core (the ratio between the unresolved core and the total VLBA emission).  \\

\item No correlation is found between the intrinsic radio luminosity at 15 GHz and the Lorentz factor of the jet for 48 quasars. A larger number of the MOJAVE-1 AGN with measured Doppler factors are needed to confirm this result.  \\

\item We find that the radio-loudness of quasars increases with increasing the redshifts. This positive correlation is due to the lack of compact AGN with small radio-loudness at high redshifts. This effect is interpreted as a tendency for strong AGN detected at high redshifts to have a high radio-loudness. This can happen if the Doppler factor of the jet in the radio regime is higher than that in the optical.

\end{enumerate}

\begin{acknowledgements}
We thank Julia Riley for useful discussions and acknowledge the contributions of Yuri Y. Kovalev,  Talvikki Hovatta, Alexander B. Pushkarev, and the rest of the MOJAVE team. Special thanks are given to the technical staff and night assistant of the OAN-SPM and OAGH. TGA acknowledges support by DFG-SPP project under grant 566960. JT acknowledges financial support from CONACYT No. 176551 scholarship. This work is supported by CONACYT basic research grants No. 39560-F, 48484 and 54480 (M\'exico). This research has made use of (1) the NASA/IPAC Extragalactic Database (NED) which is operated by the Jet Propulsion Laboratory, California Institute of Technology, under contract with the National Aeronautics and Space
Administration, (2) the USNO-B catalog (Monet et al. 2003) and MAPS Catalog of POSS~I supported by the University of Minnesota, (3) the MAPS Catalog of POSS I supported by the University of Minnesota (the APS databases can be accessed at http://aps.umn.edu/),
(4) the database CATS (Verkhodanov et al. 1997) of the Special Astrophysical Observatory, and (5) the MOJAVE database that is maintained by the MOJAVE team \citep{lister09a}. The MOJAVE project is supported under National Science Foundation grant 0807860-AST and NASA-Fermi grant NNX08AV67G. The APS databases can be accessed at http://aps.umn.edu/. The National Radio Astronomy Observatory is a facility of the National Science Foundation operated
under cooperative agreement by Associated Universities, Inc. 
\end{acknowledgements}

\bibliographystyle{aa} 

\begin{small}
\begin{longtable}{lllcrcccrcccrc}
\caption{\label{tab:1}  Kendall's $\tau$ correlation analysis between parameters of the jet, extended radio structure for all AGN from the MOJAVE-1 sample, quasars and BL Lacs. A1 and A2 are the independent variables for which the Kendall's $\tau$ correlation analysis is performed, and A3 is the dependent variable (if exists then the partial Kendall's $\tau$ correlation analysis is applied to A1 and A2),  $N$ is the number of sources included in the statistical analysis, $\tau$ is the correlation coefficient, and $P$ is the probability of a chance correlation. The correlations are considered to be significant (marked bold face) for the samples of all 135 sources and 95 quasars if the significance level $P<2\times10^{-2}$ (or confidence level $>98$\,\%), and for the sample of 18 BL Lacs if the significance level $P<5\times10^{-2}$ ($>95$\,\%). }\\
\hline\hline

&  &  &   \multicolumn{3}{c}{All} & & \multicolumn{3}{c}{Quasars} & & \multicolumn{3}{c}{BL Lac}\\
\cline{4-6}\cline{8-10}\cline{12-14} \smallskip
A1 & A2 & A3  & $N$ & $\tau$ & $P$ & & $N$ & $\tau$ & $P$ && $N$ & $\tau$ & $P$ \\

\hline

\endfirsthead
\caption{continued.}\\

\hline\hline

&  &  &   \multicolumn{3}{c}{All} & & \multicolumn{3}{c}{Quasars} & & \multicolumn{3}{c}{BL Lac}\\
\cline{4-6}\cline{8-10}\cline{12-14}
A1 & A2 & A3  & $N$ & $\tau$ & $P$ & & $N$ & $\tau$ & $P$ & & $N$ & $\tau$ & $P$ \\

\hline
\endhead
\hline
\endfoot
	$ L_{\rm VLBA}$	&	$ L_{\rm un}$	&	z	&	128	&	0.763	&	\textbf{	1.57E-16	}	&	&	101	&	0.771	&	\textbf{	9.59E-20	}	&	&	19	&	0.789	&	\textbf{	4.50E-02	}	\\
	$ L_{\rm VLBA}$	&	$ L_{\rm jet}$	&	z	&	128	&	0.471	&	\textbf{	5.02E-11	}	&	&	101	&	0.466	&	\textbf{	1.65E-11	}	&	&	19	&	0.205	&		2.16E-01		\\
	$\beta_{\rm a}$	&	$ L_{\rm VLBA}$	&	z	&	119	&	0.062	&		2.24E-01		&	&	95	&	 $-$0.011	&		8.50E-01		&	&	16	&	0.025	&		8.37E-01		\\
	$\beta_{\rm a}$	&	$ L_{\rm un}$	&	z	&	119	&	0.056	&		2.22E-01		&	&	95	&	 $-$0.004	&		9.40E-01		&	&	16	&	0.078	&		5.10E-01		\\
	$\beta_{\rm a}$	&	$ L_{\rm jet}$	&	z	&	119	&	0.104	&		6.65E-02		&	&	95	&	0.026	&		6.96E-01		&	&	16	&	0.165	&		1.70E-01		\\
	$\beta_{\rm a}$	&	$C_{\rm un}$	&	-	&	119	&	0.083	&		1.81E-01		&	&	95	&	0.055	&		4.31E-01		&	&	16	&	 $-$0.050	&		7.87E-01		\\
	$\beta_{\rm a}$	&	$C_{\rm jet}$	&	-	&	119	&	 $-$0.068	&		2.75E-01		&	&	95	&	0.027	&		6.97E-01		&	&	16	&	 $-$0.050	&		7.87E-01		\\
	$\beta_{\rm a}$	&	$C_{\rm VLBA}$	&	-	&	119	&	0.036	&		5.60E-01		&	&	95	&	 $-$0.024	&		7.28E-01		&	&	16	&	0.133	&		4.71E-01		\\
	$\beta_{\rm a}$	&	$S_{\rm VLBA}/F_{\rm B}$	&	-	&	118	&	0.038	&		5.44E-01		&	&	95	&	 $-$0.038	&		5.82E-01		&	&	15	&	 $-$0.295	&		1.25E-01		\\
	$\theta$	&	$ L_{\rm VLBA}$	&	-	&	65	&	 $-$0.467	&	\textbf{	3.74E-08	}	&	&	48	&	 $-$0.372	&	\textbf{	1.89E-04	}	&	&	12	&	 $-$0.212	&		3.37E-01		\\
	$\theta$	&	$ L_{\rm un}$	&	-	&	65	&	 $-$0.451	&	\textbf{	1.09E-07	}	&	&	48	&	 $-$0.346	&	\textbf{	5.28E-04	}	&	&	12	&	 $-$0.152	&		4.93E-01		\\
	$\theta$	&	$ L_{\rm jet}$	&	-	&	65	&	 $-$0.421	&	\textbf{	7.07E-07	}	&	&	48	&	 $-$0.317	&	\textbf{	1.46E-03	}	&	&	12	&	 $-$0.030	&		8.91E-01		\\
	$\theta$	&	$C_{\rm un}$	&	-	&	65	&	0.050	&		5.56E-01		&	&	48	&	0.089	&		3.74E-01		&	&	12	&	 $-$0.091	&		6.81E-01		\\
	$\theta$	&	$C_{\rm jet}$	&	-	&	65	&	0.142	&		9.38E-02		&	&	48	&	0.064	&		5.22E-01		&	&	12	&	0.364	&		9.98E-02		\\
	$\theta$	&	$C_{\rm VLBA}$	&	-	&	65	&	 $-$0.170	&	\textbf{	4.51E-02	}	&	&	48	&	 $-$0.133	&		1.82E-01		&	&	12	&	 $-$0.242	&		2.73E-01		\\
	$\theta$	&	$S_{\rm VLBA}/F_{\rm B}$	&	-	&	64	&	 $-$0.276	&	\textbf{	1.28E-03	}	&	&	48	&	 $-$0.183	&		6.71E-02		&	&	11	&	 $-$0.273	&		2.43E-01		\\
	$Q_{\rm j}$	&	$ L_{\rm VLBA}$	&	z	&	128	&	0.354	&	\textbf{	1.30E-07	}	&	&	101	&	0.341	&	\textbf{	4.77E-07	}	&	&	19	&	0.214	&		2.16E-01		\\
	$Q_{\rm j}$	&	$ L_{\rm un}$	&	z	&	128	&	0.277	&	\textbf{	1.08E-06	}	&	&	101	&	0.264	&	\textbf{	9.54E-06	}	&	&	19	&	0.195	&		1.79E-01		\\
	$Q_{\rm j}$	&	$ L_{\rm jet}$	&	z	&	128	&	0.400	&	\textbf{	7.61E-10	}	&	&	101	&	0.414	&	\textbf{	2.16E-10	}	&	&	19	&	0.097	&		4.65E-01		\\
	$Q_{\rm j}$	&	$ L_{\rm VLBA,int}$	&	z	&	70	&	0.204	&	\textbf{	2.87E-03	}	&	&	51	&	0.205	&	\textbf{	1.58E-02	}	&	&	14	&	0.123	&		4.93E-01		\\
	$Q_{\rm j}$	&	$ L_{\rm un,int}$	&	z	&	70	&	0.236	&	\textbf{	1.29E-03	}	&	&	51	&	0.237	&	\textbf{	8.91E-03	}	&	&	14	&	0.147	&		4.17E-01		\\
	$Q_{\rm j}$	&	$ L_{\rm jet,int}$	&	z	&	70	&	0.201	&	\textbf{	5.12E-03	}	&	&	51	&	0.244	&	\textbf{	6.38E-03	}	&	&	14	&	 $-$0.014	&		9.48E-01		\\
	$Q_{\rm j}$	&	$C_{\rm un}$	&	-	&	128	&	 $-$0.174	&	\textbf{	3.66E-03	}	&	&	101	&	 $-$0.261	&	\textbf{	1.13E-04	}	&	&	19	&	0.333	&	\textbf{	4.61E-02	}	\\
	$Q_{\rm j}$	&	$C_{\rm jet}$	&	-	&	128	&	0.021	&		7.29E-01		&	&	101	&	0.155	&	\textbf{	2.14E-02	}	&	&	19	&	 $-$0.216	&		1.96E-01		\\
	$Q_{\rm j}$	&	$C_{\rm VLBA}$	&	-	&	128	&	 $-$0.122	&	\textbf{	4.16E-02	}	&	&	101	&	 $-$0.229	&	\textbf{	6.84E-04	}	&	&	19	&	 $-$0.088	&		6.00E-01		\\
	$Q_{\rm j}$	&	$S_{\rm VLBA}/F_{\rm B}$	&	-	&	125	&	0.249	&	\textbf{	3.88E-05	}	&	&	99	&	0.177	&	\textbf{	9.24E-03	}	&	&	18	&	0.320	&		6.35E-02		\\
	$Q_{\rm j}$	&	$\beta_{\rm a}$	&	z	&	119	&	0.187	&	\textbf{	6.92E-04	}	&	&	95	&	0.181	&	\textbf{	1.07E-03	}	&	&	16	&	 $-$0.083	&		4.25E-01		\\
	$Q_{\rm j}$	&	$\gamma$	&	-	&	65	&	0.291	&	\textbf{	6.02E-04	}	&	&	48	&	0.046	&		6.44E-01		&	&	12	&	0.333	&		1.31E-01		\\
	$Q_{\rm j}$	&	$\theta$	&	-	&	65	&	 $-$0.312	&	\textbf{	2.44E-04	}	&	&	48	&	 $-$0.161	&		1.06E-01		&	&	12	&	 $-$0.061	&		7.84E-01		\\
	$ L_{\rm VLBA,int}$	&	$\gamma$	&	-	&	65	&	0.037	&		6.67E-01		&	&	48	&	 $-$0.190	&		5.72E-02		&	&	12	&	0.273	&		2.17E-01		\\
	$ L_{\rm un,int}$	&	$\gamma$	&	-	&	65	&	0.059	&		4.90E-01		&	&	48	&	 $-$0.168	&		9.13E-02		&	&	12	&	0.273	&		2.17E-01		\\
	$ L_{\rm jet,int}$	&	$\gamma$	&	-	&	65	&	 $-$0.053	&		5.33E-01		&	&	48	&	 $-$0.195	&		5.05E-02		&	&	12	&	 $-$0.061	&		7.84E-01		\\
	$D_{\rm var}$	&	$C_{\rm un}$	&	-	&	72	&	0.024	&		7.66E-01		&	&	51	&	0.050	&		6.03E-01		&	&	16	&	0.042	&		8.21E-01		\\
	$D_{\rm var}$	&	$C_{\rm VLBA}$	&	-	&	72	&	0.178	&	\textbf{	2.66E-02	}	&	&	51	&	0.196	&	\textbf{	4.21E-02	}	&	&	16	&	0.276	&		1.36E-01		\\
	$D_{\rm var}$	&	$C_{\rm jet}$	&	-	&	72	&	 $-$0.258	&	\textbf{	1.37E-03	}	&	&	51	&	 $-$0.129	&		1.82E-01		&	&	16	&	 $-$0.460	&	\textbf{	1.29E-02	}	\\
	$D_{\rm var}$	&	z	&	-	&	70	&	0.381	&	\textbf{	3.13E-06	}	&	&	51	&	0.163	&		9.22E-02		&	&	14	&	0.278	&		1.66E-01		\\
	$S_{\rm VLBA}/F_{\rm B}$	&	z	&	-	&	125	&	0.327	&	\textbf{	6.51E-08	}	&	&	99	&	0.255	&	\textbf{	1.82E-04	}	&	&	18	&	0.242	&		1.61E-01		\\
	$\gamma$	&	z	&	-	&	65	&	0.189	&	\textbf{	2.61E-02	}	&	&	48	&	 $-$0.134	&		1.79E-01		&	&	12	&	0.364	&		9.98E-02		\\
	$\theta$	&	z	&	-	&	65	&	 $-$0.382	&	\textbf{	6.74E-06	}	&	&	48	&	 $-$0.244	&	\textbf{	1.45E-02	}	&	&	12	&	0.121	&		5.83E-01		\\

\end{longtable}
\end{small}


\begin{small}

\begin{longtable}{lllcrcccrcccrc}
\caption{\label{tab:2} Kendall's $\tau$ correlation analysis between radio parameters of the jet and optical nuclear luminosity at 5100\,\AA. }\\
\hline\hline

&  &  &   \multicolumn{3}{c}{All} & & \multicolumn{3}{c}{Quasars} & & \multicolumn{3}{c}{BL Lac}\\
\cline{4-6}\cline{8-10}\cline{12-14} \smallskip
A1 & A2 & A3  & $N$ & $\tau$ & $P$ & & $N$ & $\tau$ & $P$ & & $N$ & $\tau$ & $P$ \\

\hline

\endfirsthead
\caption{continued.}\\

\hline\hline

&  &  &   \multicolumn{3}{c}{All} & & \multicolumn{3}{c}{Quasars} & & \multicolumn{3}{c}{BL Lac}\\
\cline{4-6}\cline{8-10}\cline{12-14}
A1 & A2 & A3  & N & $\tau$ & P & & N & $\tau$ & P && N & $\tau$ & P \\

\hline
\endhead
\hline
\endfoot
	$L_{5100}$	&	$ L_{\rm VLBA}$	&	z	&	125	&	0.252	&	\textbf{	2.05E-07	}	&	&	99	&	0.260	&	\textbf{	8.33E-07	}	&	&	18	&	0.038	&		7.54E-01		\\
	$L_{5100}$	&	$ L_{\rm un}$	&	z	&	125	&	0.220	&	\textbf{	1.73E-06	}	&	&	99	&	0.217	&	\textbf{	1.37E-05	}	&	&	18	&	0.021	&		8.54E-01		\\
	$L_{5100}$	&	$ L_{\rm jet}$	&	z	&	125	&	0.285	&	\textbf{	4.84E-09	}	&	&	99	&	0.254	&	\textbf{	7.56E-06	}	&	&	18	&	0.407	&	\textbf{	1.30E-03	}	\\
	$L_{5100}$	&	$\beta_{\rm a}$	&	z	&	118	&	0.035	&		5.28E-01		&	&	95	&	 $-$0.056	&		3.75E-01		&	&	15	&	0.402	&	\textbf{	4.20E-02	}	\\
	$L_{5100}$	&	$\gamma$	&	-	&	64	&	0.148	&		8.43E-02		&	&	48	&	 $-$0.147	&		1.40E-01		&	&	11	&	0.600	&	\textbf{	1.02E-02	}	\\
	$L_{5100}$	&	$\theta$	&	-	&	64	&	 $-$0.350	&	\textbf{	4.31E-05	}	&	&	48	&	 $-$0.309	&	\textbf{	1.98E-03	}	&	&	11	&	 $-$0.018	&		9.38E-01		\\
	$L_{5100}$	&	$Q_{\rm j}$	&	z	&	125	&	0.187	&	\textbf{	4.39E-05	}	&	&	99	&	0.179	&	\textbf{	9.89E-04	}	&	&	18	&	 $-$0.066	&		5.80E-01		\\
	$L_{5100}$	&	$C_{\rm un}$	&	-	&	125	&	 $-$0.087	&		1.51E-01		&	&	99	&	 $-$0.177	&	\textbf{	9.41E-03	}	&	&	18	&	 $-$0.033	&		8.50E-01		\\
	$L_{5100}$	&	$C_{\rm jet}$	&	-	&	125	&	 $-$0.070	&		2.46E-01		&	&	99	&	0.025	&		7.10E-01		&	&	18	&	0.085	&		6.22E-01		\\
	$L_{5100}$	&	$C_{\rm VLBA}$	&	-	&	125	&	0.048	&		4.26E-01		&	&	99	&	0.003	&		9.61E-01		&	&	18	&	 $-$0.176	&		3.06E-01		\\
	$L_{5100}$	&	$S_{\rm VLBA}/F_{\rm B}$	&	-	&	125	&	 $-$0.004	&		9.52E-01		&	&	99	&	 $-$0.088	&		1.95E-01		&	&	18	&	 $-$0.359	&	\textbf{	3.72E-02	}	\\
\end{longtable}

\end{small}

\begin{normalsize}

\begin{longtable}{ccccccc}

\caption{\label{tab:3} Optical and radio parameters of the 135 AGN from the MOJAVE-1 sample. Column (1) is the name of the object; (2) is the spectral type: quasars (Q), BL Lacs (BL), radio galaxies (G), and unclassified objects (U); (3) is the logarithm of the nuclear optical flux and its error in units of Janskys; (4) is the logarithm of the nuclear optical luminosity at 5100\, \AA\, in units of W\,Hz$^{-1}$; (5) is the  flux density at 151\,MHz in units of Janskys; (6) is the logarithm of the power of the jet in units of Watts; (7) is the apparent speed  of the jet and its error in units of the speed of the light.} \\

\hline\hline\\ [-2ex]

Name & Sp. Type  & log $F_{\rm B, \,corr}$ & log $L_{5100}$ & $S_{\rm 151}$ & log $Q_{\rm j}$ & $\beta_{\rm a}$\\

(1) & (2) &(3) &(4) &(5) &(6) &(7) \\

\hline\\ [-2ex]

\endfirsthead
\caption{continued.}\\

\hline\hline\\ [-2ex]

Name & Sp. Type  & log $F_{\rm B, \,corr}$ & log $L_{5100}$ & $S_{\rm 151}$ & log $Q_{\rm j}$ & $\beta_{\rm a}$\\

(1) & (2) &(3) &(4) &(5) &(6) &(7) \\

\hline\\ [-2ex]
\endhead
\hline
\endfoot

0003$-$066	&	BL	&	$-$4.73	$\,\pm\,$	0.09	&	21.93	&	2.04	&	37.34	&	0.89	$\,\pm\,$	0.17	\\
0007+106	&	G	&	$-$2.38	$\,\pm\,$	0.11	&	23.06	&	0.05	&	34.38	&	0.99	$\,\pm\,$	0.09	\\
0016+731	&	Q	&	$-$4.37	$\,\pm\,$	0.14	&	24.43	&	0.24	&	38.14	&	6.66	$\,\pm\,$	0.38	\\
0048$-$097	&	BL	&		\ldots		&	\ldots	&	1.14	&	\ldots	&		\ldots		\\
0059+581	&	Q	&	$-$4.33	$\,\pm\,$	0.18	&	23.71	&	0.88	&	37.61	&	11.12	$\,\pm\,$	0.85	\\
0106+013	&	Q	&	$-$4.42	$\,\pm\,$	0.06	&	24.14	&	5.06	&	39.63	&	26.12	$\,\pm\,$	3.86	\\
0109+224	&	BL	&	$-$3.14	$\,\pm\,$	0.10	&	23.25	&	0.34	&	36.30	&		\ldots		\\
0119+115	&	Q	&	$-$4.59	$\,\pm\,$	0.08	&	22.57	&	2.52	&	37.94	&	17.17	$\,\pm\,$	0.67	\\
0133+476	&	Q	&	$-$3.82	$\,\pm\,$	0.16	&	23.95	&	1.47	&	38.14	&	12.96	$\,\pm\,$	2.52	\\
0202+149	&	Q	&	$-$5.52	$\,\pm\,$	0.21	&	21.32	&	6.22	&	37.98	&	6.46	$\,\pm\,$	1.29	\\
0202+319	&	Q	&	$-$3.59	$\,\pm\,$	0.07	&	24.62	&	0.65	&	38.36	&	2.99	$\,\pm\,$	0.95	\\
0212+735	&	Q	&	$-$4.79	$\,\pm\,$	0.14	&	24.92	&	1.11	&	39.10	&	7.52	$\,\pm\,$	0.27	\\
0215+015	&	Q	&	$-$4.88	$\,\pm\,$	0.15	&	23.46	&	1.47	&	38.88	&	33.75	$\,\pm\,$	2.10	\\
0224+671	&	Q	&	$-$4.78	$\,\pm\,$	0.19	&	23.73	&	2.37	&	37.83	&	11.69	$\,\pm\,$	0.47	\\
0234+285	&	Q	&	$-$4.43	$\,\pm\,$	0.07	&	23.71	&	2.33	&	38.70	&	12.18	$\,\pm\,$	0.84	\\
0235+164	&	BL	&	$-$4.82	$\,\pm\,$	0.08	&	22.95	&	1.07	&	38.10	&		\ldots		\\
0238$-$084	&	G	&	$-$2.17	$\,\pm\,$	0.14	&	20.61	&	2.51	&	33.81	&	0.35	$\,\pm\,$	0.01	\\
0300+470	&	BL	&		\ldots		&	\ldots	&	1.35	&	\ldots	&		\ldots		\\
0316+413	&	G	&	$-$2.11	$\,\pm\,$	0.23	&	21.99	&	57.94	&	36.12	&	0.32	$\,\pm\,$	0.06	\\
0333+321	&	Q	&	$-$3.97	$\,\pm\,$	0.07	&	25.08	&	3.06	&	38.87	&	12.66	$\,\pm\,$	0.18	\\
0336$-$019	&	Q	&	$-$3.77	$\,\pm\,$	0.07	&	23.90	&	1.58	&	38.16	&	22.33	$\,\pm\,$	3.66	\\
0403$-$132	&	Q	&	$-$3.91	$\,\pm\,$	0.08	&	23.29	&	9.84	&	38.54	&	19.77	$\,\pm\,$	0.87	\\
0415+379	&	G	&	$-$4.36	$\,\pm\,$	0.25	&	22.76	&	85.31	&	37.13	&	5.96	$\,\pm\,$	0.10	\\
0420$-$014	&	Q	&	$-$3.91	$\,\pm\,$	0.08	&	23.90	&	1.85	&	38.31	&	7.33	$\,\pm\,$	0.98	\\
0422+004	&	BL	&	$-$3.47	$\,\pm\,$	0.10	&	23.17	&	2.33	&	37.29	&		\ldots		\\
0430+052	&	G	&	$-$3.18	$\,\pm\,$	0.12	&	21.65	&	9.51	&	35.84	&	5.46	$\,\pm\,$	0.16	\\
0446+112	&	U	&		\ldots		&	\ldots	&	1.20	&	\ldots	&		\ldots		\\
0458$-$020	&	Q	&	$-$4.52	$\,\pm\,$	0.06	&	24.20	&	4.19	&	39.64	&	16.26	$\,\pm\,$	0.79	\\
0528+134	&	Q	&	$-$4.77	$\,\pm\,$	0.14	&	24.93	&	0.62	&	38.71	&	18.92	$\,\pm\,$	0.42	\\
0529+075	&	Q	&	$-$4.63	$\,\pm\,$	0.15	&	23.78	&	1.94	&	38.67	&	12.57	$\,\pm\,$	1.47	\\
0529+483	&	Q	&	$-$4.90	$\,\pm\,$	0.16	&	23.57	&	0.79	&	38.19	&	19.66	$\,\pm\,$	2.97	\\
0552+398	&	Q	&	$-$4.10	$\,\pm\,$	0.13	&	25.16	&	0.16	&	38.27	&	0.36	$\,\pm\,$	0.09	\\
0605$-$085	&	Q	&	$-$4.36	$\,\pm\,$	0.25	&	24.07	&	3.37	&	38.52	&	16.74	$\,\pm\,$	0.42	\\
0607$-$157	&	Q	&	$-$4.20	$\,\pm\,$	0.20	&	22.71	&	1.60	&	37.17	&	3.98	$\,\pm\,$	1.08	\\
0642+449	&	Q	&	$-$4.67	$\,\pm\,$	0.05	&	24.51	&	0.32	&	38.94	&	0.74	$\,\pm\,$	0.11	\\
0648$-$165	&	U	&		\ldots		&	\ldots	&	2.22	&	\ldots	&		\ldots		\\
0716+714	&	BL	&		\ldots		&	\ldots	&	2.37	&	37.29	&	10.17	$\,\pm\,$	0.35	\\
0727$-$115	&	Q	&		\ldots		&	\ldots	&	2.00	&	38.93	&		\ldots		\\
0730+504	&	Q	&	$-$4.54	$\,\pm\,$	0.08	&	22.91	&	0.88	&	37.73	&	14.08	$\,\pm\,$	4.18	\\
0735+178	&	BL	&		\ldots		&	\ldots	&	2.61	&	\ldots	&		\ldots		\\
0736+017	&	Q	&	$-$3.68	$\,\pm\,$	0.22	&	22.52	&	2.96	&	36.91	&	14.62	$\,\pm\,$	0.95	\\
0738+313	&	Q	&	$-$3.41	$\,\pm\,$	0.08	&	23.88	&	0.98	&	37.64	&	10.79	$\,\pm\,$	1.13	\\
0742+103	&	Q	&		\ldots		&	\ldots	&	2.13	&	39.49	&		\ldots		\\
0748+126	&	Q	&	$-$3.88	$\,\pm\,$	0.16	&	23.76	&	1.82	&	38.27	&	18.33	$\,\pm\,$	0.81	\\
0754+100	&	BL	&	$-$3.04	$\,\pm\,$	0.10	&	23.34	&	0.83	&	36.68	&	14.57	$\,\pm\,$	1.22	\\
0804+499	&	Q	&	$-$3.02	$\,\pm\,$	0.07	&	25.16	&	0.39	&	38.11	&	1.81	$\,\pm\,$	0.34	\\
0805$-$077	&	Q	&	$-$4.09	$\,\pm\,$	0.14	&	24.48	&	3.93	&	39.38	&	49.95	$\,\pm\,$	2.03	\\
0808+019	&	BL	&	$-$4.09	$\,\pm\,$	0.15	&	23.82	&	0.62	&	38.08	&	12.92	$\,\pm\,$	0.81	\\
0814+425	&	BL	&	$-$4.68	$\,\pm\,$	0.10	&	21.67	&	1.91	&	36.97	&	1.73	$\,\pm\,$	0.30	\\
0823+033	&	BL	&	$-$3.15	$\,\pm\,$	0.09	&	23.91	&	0.77	&	37.30	&	17.90	$\,\pm\,$	0.82	\\
0827+243	&	Q	&	$-$3.72	$\,\pm\,$	0.08	&	23.98	&	0.74	&	37.94	&	21.91	$\,\pm\,$	1.85	\\
0829+046	&	BL	&	$-$2.80	$\,\pm\,$	0.10	&	23.21	&	1.71	&	36.59	&	10.24	$\,\pm\,$	0.39	\\
0836+710	&	Q	&	$-$3.73	$\,\pm\,$	0.06	&	24.88	&	4.70	&	39.66	&	25.00	$\,\pm\,$	0.95	\\
0838+133 	&	Q	&	$-$3.98	$\,\pm\,$	0.08	&	23.47	&	13.07	&	38.84	&	12.96	$\,\pm\,$	1.16	\\
0851+202	&	BL	&	$-$2.78	$\,\pm\,$	0.10	&	23.74	&	0.71	&	36.76	&	5.26	$\,\pm\,$	0.40	\\
0906+015	&	Q	&	$-$3.64	$\,\pm\,$	0.07	&	24.15	&	1.15	&	38.22	&	20.57	$\,\pm\,$	0.85	\\
0917+624	&	Q	&	$-$4.77	$\,\pm\,$	0.09	&	23.42	&	1.14	&	38.59	&	12.05	$\,\pm\,$	1.62	\\
0923+392	&	Q	&	$-$3.56	$\,\pm\,$	0.17	&	23.80	&	6.75	&	38.58	&	0.59	$\,\pm\,$	0.16	\\
0945+408	&	Q	&	$-$3.96	$\,\pm\,$	0.07	&	24.02	&	3.16	&	38.87	&	18.46	$\,\pm\,$	0.86	\\
0955+476	&	Q	&	$-$4.25	$\,\pm\,$	0.07	&	24.16	&	0.37	&	38.38	&	2.45	$\,\pm\,$	0.24	\\
1036+054	&	Q	&	$-$4.78	$\,\pm\,$	0.09	&	22.18	&	1.00	&	37.35	&	6.02	$\,\pm\,$	0.88	\\
1038+064	&	Q	&	$-$3.56	$\,\pm\,$	0.07	&	24.45	&	1.83	&	38.65	&	11.78	$\,\pm\,$	0.99	\\
1045$-$188	&	Q	&	$-$3.93	$\,\pm\,$	0.08	&	23.29	&	4.97	&	38.28	&	8.60	$\,\pm\,$	0.69	\\
1055+018	&	Q	&	$-$4.02	$\,\pm\,$	0.08	&	23.60	&	5.56	&	38.76	&	8.05	$\,\pm\,$	1.44	\\
1124$-$186	&	Q	&	$-$4.57	$\,\pm\,$	0.07	&	23.26	&	0.66	&	38.01	&		\ldots		\\
1127$-$145	&	Q	&	$-$3.65	$\,\pm\,$	0.07	&	24.31	&	4.04	&	38.92	&	14.08	$\,\pm\,$	0.59	\\
1150+812	&	Q	&	$-$4.68	$\,\pm\,$	0.07	&	23.39	&	1.90	&	38.65	&	5.38	$\,\pm\,$	0.92	\\
1156+295	&	Q	&	$-$3.67	$\,\pm\,$	0.08	&	23.73	&	4.28	&	38.43	&	24.88	$\,\pm\,$	1.84	\\
1213$-$172	&	U	&		\ldots		&	\ldots	&	4.29	&	\ldots	&		\ldots		\\
1219+044	&	Q	&	$-$3.81	$\,\pm\,$	0.07	&	23.90	&	2.47	&	38.49	&	2.34	$\,\pm\,$	0.41	\\
1222+216	&	Q	&	$-$3.94	$\,\pm\,$	0.09	&	22.93	&	7.66	&	38.14	&	21.17	$\,\pm\,$	1.90	\\
1226+023	&	Q	&	$-$3.45	$\,\pm\,$	0.10	&	22.43	&	97.95	&	38.25	&	13.62	$\,\pm\,$	0.43	\\
1228+126	&	G	&	$-$1.54	$\,\pm\,$	0.12	&	21.04	&	1180.32	&	36.34	&	0.03	$\,\pm\,$	0.00	\\
1253$-$055	&	Q	&	$-$3.33	$\,\pm\,$	0.09	&	23.77	&	22.08	&	38.82	&	20.68	$\,\pm\,$	0.79	\\
1308+326	&	Q	&	$-$4.71	$\,\pm\,$	0.07	&	23.02	&	1.55	&	38.32	&	20.80	$\,\pm\,$	0.68	\\
1324+224	&	Q	&	$-$4.09	$\,\pm\,$	0.07	&	24.01	&	0.49	&	38.19	&		\ldots		\\
1334$-$127	&	Q	&	$-$4.26	$\,\pm\,$	0.09	&	22.91	&	3.66	&	38.05	&	10.31	$\,\pm\,$	0.96	\\
1413+135	&	BL	&	$-$5.17	$\,\pm\,$	0.22	&	21.13	&	4.65	&	37.36	&	1.82	$\,\pm\,$	0.17	\\
1417+385	&	Q	&	$-$4.96	$\,\pm\,$	0.07	&	23.42	&	0.47	&	38.46	&	15.25	$\,\pm\,$	2.96	\\
1458+718	&	Q	&	$-$3.69	$\,\pm\,$	0.08	&	23.95	&	27.16	&	39.46	&	7.03	$\,\pm\,$	2.04	\\
1502+106	&	Q	&	$-$4.63	$\,\pm\,$	0.06	&	23.79	&	0.91	&	38.75	&	14.58	$\,\pm\,$	1.17	\\
1504$-$166	&	Q	&	$-$4.80	$\,\pm\,$	0.08	&	22.91	&	2.84	&	38.45	&	4.02	$\,\pm\,$	0.14	\\
1510$-$089	&	Q	&	$-$3.31	$\,\pm\,$	0.09	&	23.47	&	6.05	&	37.85	&	20.33	$\,\pm\,$	1.16	\\
1538+149	&	BL	&	$-$3.29	$\,\pm\,$	0.08	&	23.97	&	5.35	&	38.33	&	8.76	$\,\pm\,$	0.95	\\
1546+027	&	Q	&	$-$3.96	$\,\pm\,$	0.09	&	22.98	&	1.10	&	37.25	&	12.17	$\,\pm\,$	1.27	\\
1548+056	&	Q	&	$-$4.15	$\,\pm\,$	0.07	&	24.04	&	3.16	&	39.01	&	11.46	$\,\pm\,$	1.67	\\
1606+106	&	Q	&	$-$4.12	$\,\pm\,$	0.07	&	23.91	&	3.47	&	38.89	&	17.81	$\,\pm\,$	1.09	\\
1611+343	&	Q	&	$-$3.99	$\,\pm\,$	0.07	&	24.12	&	2.51	&	38.89	&	5.66	$\,\pm\,$	0.59	\\
1633+382	&	Q	&	$-$4.20	$\,\pm\,$	0.07	&	24.16	&	2.57	&	39.18	&	29.09	$\,\pm\,$	1.62	\\
1637+574	&	Q	&	$-$3.59	$\,\pm\,$	0.08	&	23.84	&	1.93	&	38.12	&	10.62	$\,\pm\,$	1.26	\\
1638+398	&	Q	&	$-$4.30	$\,\pm\,$	0.07	&	23.98	&	0.78	&	38.57	&	12.14	$\,\pm\,$	1.54	\\
1641+399	&	Q	&	$-$3.33	$\,\pm\,$	0.08	&	23.85	&	12.39	&	38.68	&	19.34	$\,\pm\,$	0.52	\\
1655+077	&	Q	&	$-$4.74	$\,\pm\,$	0.18	&	22.69	&	2.62	&	38.05	&	14.49	$\,\pm\,$	1.15	\\
1726+455	&	Q	&	$-$4.31	$\,\pm\,$	0.08	&	23.08	&	0.47	&	37.45	&	1.82	$\,\pm\,$	0.44	\\
1730$-$130	&	Q	&	$-$5.18	$\,\pm\,$	0.17	&	23.16	&	6.87	&	38.86	&	35.61	$\,\pm\,$	2.13	\\
1739+522	&	Q	&	$-$4.17	$\,\pm\,$	0.07	&	23.94	&	1.02	&	38.49	&		\ldots		\\
1741$-$038	&	Q	&	$-$4.55	$\,\pm\,$	0.16	&	24.04	&	6.40	&	39.00	&		\ldots		\\
1749+096	&	BL	&	$-$4.12	$\,\pm\,$	0.10	&	22.67	&	7.60	&	37.84	&	6.91	$\,\pm\,$	0.79	\\
1751+288	&	Q	&	$-$5.04	$\,\pm\,$	0.07	&	22.88	&	0.63	&	38.05	&	3.06	$\,\pm\,$	0.74	\\
1758+388	&	Q	&	$-$4.07	$\,\pm\,$	0.06	&	24.48	&	0.21	&	38.24	&	2.35	$\,\pm\,$	0.34	\\
1800+440	&	Q	&	$-$3.67	$\,\pm\,$	0.08	&	23.70	&	2.80	&	38.15	&	15.45	$\,\pm\,$	0.49	\\
1803+784	&	BL	&	$-$3.11	$\,\pm\,$	0.08	&	24.28	&	2.09	&	38.05	&	8.99	$\,\pm\,$	2.52	\\
1807+698	&	BL	&	$-$3.60	$\,\pm\,$	0.11	&	21.25	&	9.36	&	36.21	&	0.11	$\,\pm\,$	0.02	\\
1823+568	&	BL	&	$-$4.23	$\,\pm\,$	0.08	&	23.13	&	3.79	&	38.28	&	9.42	$\,\pm\,$	1.94	\\
1828+487	&	Q	&	$-$3.74	$\,\pm\,$	0.08	&	23.70	&	79.43	&	39.64	&	13.68	$\,\pm\,$	0.39	\\
1849+670	&	Q	&	$-$3.39	$\,\pm\,$	0.08	&	23.96	&	1.16	&	37.75	&	30.70	$\,\pm\,$	1.51	\\
1928+738	&	Q	&	$-$3.16	$\,\pm\,$	0.10	&	23.50	&	5.45	&	37.63	&	7.32	$\,\pm\,$	1.10	\\
1936$-$155	&	Q	&	$-$4.57	$\,\pm\,$	0.15	&	23.92	&	0.72	&	38.53	&	2.57	$\,\pm\,$	0.73	\\
1957+405	&	G	&	$-$2.39	$\,\pm\,$	0.23	&	23.03	&	8770.01	&	39.26	&	0.22	$\,\pm\,$	0.04	\\
1958$-$179	&	Q	&	$-$3.96	$\,\pm\,$	0.08	&	23.50	&	0.98	&	37.67	&	1.90	$\,\pm\,$	0.19	\\
2005+403	&	Q	&	$-$4.58	$\,\pm\,$	0.15	&	24.73	&	0.77	&	38.61	&	4.37	$\,\pm\,$	1.61	\\
2008$-$159	&	Q	&	$-$3.81	$\,\pm\,$	0.15	&	24.29	&	1.03	&	38.33	&	7.94	$\,\pm\,$	0.94	\\
2021+317	&	U	&		\ldots		&	\ldots	&	2.84	&	\ldots	&		\ldots		\\
2021+614	&	G	&	$-$4.57	$\,\pm\,$	0.22	&	21.91	&	0.55	&	36.35	&	0.10	$\,\pm\,$	0.01	\\
2037+511	&	Q	&	$-$4.51	$\,\pm\,$	0.15	&	25.34	&	18.79	&	39.97	&	3.26	$\,\pm\,$	1.87	\\
2121+053	&	Q	&	$-$4.12	$\,\pm\,$	0.06	&	24.41	&	0.93	&	38.81	&	8.28	$\,\pm\,$	1.05	\\
2128$-$123	&	Q	&	$-$3.18	$\,\pm\,$	0.09	&	23.89	&	2.04	&	37.72	&	1.08	$\,\pm\,$	0.21	\\
2131$-$021	&	BL	&	$-$4.35	$\,\pm\,$	0.07	&	23.72	&	2.78	&	38.85	&	19.88	$\,\pm\,$	1.43	\\
2134+004	&	Q	&	$-$3.76	$\,\pm\,$	0.06	&	24.76	&	1.73	&	39.08	&	2.01	$\,\pm\,$	0.24	\\
2136+141	&	Q	&	$-$4.41	$\,\pm\,$	0.06	&	24.40	&	0.86	&	39.01	&	2.90	$\,\pm\,$	0.18	\\
2145+067	&	Q	&	$-$3.53	$\,\pm\,$	0.07	&	24.30	&	3.50	&	38.67	&	2.18	$\,\pm\,$	0.16	\\
2155$-$152	&	Q	&	$-$3.91	$\,\pm\,$	0.08	&	23.46	&	3.14	&	38.21	&	18.15	$\,\pm\,$	1.77	\\
2200+420	&	BL	&	$-$3.04	$\,\pm\,$	0.23	&	22.49	&	1.77	&	35.75	&	5.05	$\,\pm\,$	0.30	\\
2201+171	&	Q	&	$-$4.59	$\,\pm\,$	0.07	&	23.29	&	1.64	&	38.43	&	1.54	$\,\pm\,$	0.33	\\
2201+315	&	Q	&	$-$2.91	$\,\pm\,$	0.10	&	23.72	&	3.50	&	37.41	&	7.96	$\,\pm\,$	0.41	\\
2209+236	&	Q	&	$-$4.69	$\,\pm\,$	0.08	&	23.26	&	0.41	&	37.88	&	3.41	$\,\pm\,$	0.51	\\
2216$-$038	&	Q	&	$-$3.65	$\,\pm\,$	0.08	&	24.09	&	3.21	&	38.53	&	5.61	$\,\pm\,$	0.53	\\
2223$-$052	&	Q	&	$-$4.10	$\,\pm\,$	0.07	&	24.09	&	18.84	&	39.77	&	14.47	$\,\pm\,$	1.26	\\
2227$-$088	&	Q	&	$-$4.02	$\,\pm\,$	0.07	&	24.25	&	2.72	&	39.05	&	8.05	$\,\pm\,$	2.05	\\
2230+114	&	Q	&	$-$3.74	$\,\pm\,$	0.07	&	24.12	&	5.66	&	38.93	&	15.34	$\,\pm\,$	0.64	\\
2243$-$123	&	Q	&	$-$3.42	$\,\pm\,$	0.08	&	23.88	&	1.18	&	37.72	&	5.51	$\,\pm\,$	0.33	\\
2251+158	&	Q	&	$-$3.52	$\,\pm\,$	0.16	&	24.19	&	14.03	&	39.12	&	14.17	$\,\pm\,$	0.79	\\
2331+073	&	Q	&	$-$3.43	$\,\pm\,$	0.09	&	23.44	&	0.52	&	36.89	&	4.51	$\,\pm\,$	0.47	\\
2345$-$167	&	Q	&	$-$4.20	$\,\pm\,$	0.08	&	22.97	&	3.33	&	38.07	&	13.51	$\,\pm\,$	0.81	\\
2351+456	&	Q	&	$-$4.99	$\,\pm\,$	0.14	&	23.63	&	2.38	&	39.25	&	16.37	$\,\pm\,$	1.47	\\

\end{longtable}

\end{normalsize}

\end{document}